\documentclass[twocolumn,,amsmath,amssymb]{revtex4}

\pdfoutput=1

\usepackage{graphicx}
\usepackage{dcolumn}
\usepackage{bm}
\newcommand{\etal}{\emph{et al.}}
\newcommand{\be}{\begin{equation}}
\newcommand{\ee}{\end{equation}}
\newcommand{\bfig}{\begin{figure}}
\newcommand{\efig}{\end{figure}}

\begin{document}      
\title{The chiral anomaly and thermopower of Weyl fermions in the half-Heusler GdPtBi 
} 

\author{Max Hirschberger$^{1}$, Satya Kushwaha$^2$, Zhijun Wang$^{1}$, Quinn Gibson$^{2}$, Sihang Liang$^1$, Carina A. Belvin$^{1,\ddag}$, B. A. Bernevig$^1$, R. J. Cava$^2$ and N. P. Ong$^1$}
\affiliation{
$^1$Department of Physics, Princeton University, Princeton, NJ 08544\\
$^2$Department of Chemistry, Princeton University, Princeton, NJ 08544
}

\date{\today}      
\begin{abstract}
\end{abstract}
 
\maketitle      
{\bf  The Dirac and Weyl semimetals are unusual materials in which the nodes of the bulk states are protected against gap formation by crystalline symmetry~\cite{Ashvin,Kane,Bernevig,Wang1}. The chiral anomaly~\cite{Adler,Bell}, predicted to occur in both systems~\cite{Nielsen,Son,Balents,Burkov}, was recently observed as a negative longitudinal magnetoresistance (LMR) in Na$_3$Bi~\cite{Xiong} and in TaAs~\cite{Dai}. An important issue is whether Weyl physics appears in a broader class of materials. We report evidence for the chiral anomaly in the half-Heusler GdPtBi. In zero field, GdPtBi is a zero-gap semiconductor with quadratic bands~\cite{Canfield,Mong}. In a magnetic field, the Zeeman energy leads to Weyl nodes~\cite{Moon}. We have observed a large negative LMR with the field-steering properties specific to the chiral anomaly. The chiral anomaly also induces strong suppression of the thermopower. We report a detailed study of the thermoelectric response function $\alpha_{xx}$ of Weyl fermions. The scheme of creating Weyl nodes from quadratic bands suggests that the chiral anomaly may be observable in a broad class of semimetals.
}

The Dirac semimetal Na$_3$Bi exhibits two bulk Dirac cones in zero field~\cite{Wang1,Chen}. Each cone resolves into two Weyl nodes with distinct chiralities $\chi = \pm 1$. In a magnetic field $\bf B$, the Weyl nodes separate in $\bf k$ (momentum) space to act as monopole source and sink of the Berry curvature $\bf \Omega(k)$ (which acts as an effective magnetic field in $\bf k$ space~\cite{Niu}). As observed in Na$_3$Bi~\cite{Xiong}, the application of an electric field $\bf E\parallel B$ produces a negative longitudinal magnetoresistance (LMR) produced by the chiral anomaly. The recently discovered Weyl semimetals~\cite{Hasan,Bernevig2} are similar, except that the Weyl nodes are already separated at $B$ = 0 because inversion symmetry is broken~\cite{Dai}. Here we demonstrate a third route towards Weyl nodes, starting with a material with $T_d$ symmetry and displaying quadratic bands that touch~\cite{Moon}. In finite $B$, the Zeeman energy leads to band crossings and the formation of Weyl nodes. In the half-Heusler GdPtBi, this scheme results in the appearance of the chiral anomaly (in samples with the Fermi energy $E_F$ much closer to the Weyl node than in previous experiments~\cite{Canfield,Paglione,Muller}). 

\begin{figure*}[t]
\includegraphics[width=16 cm]{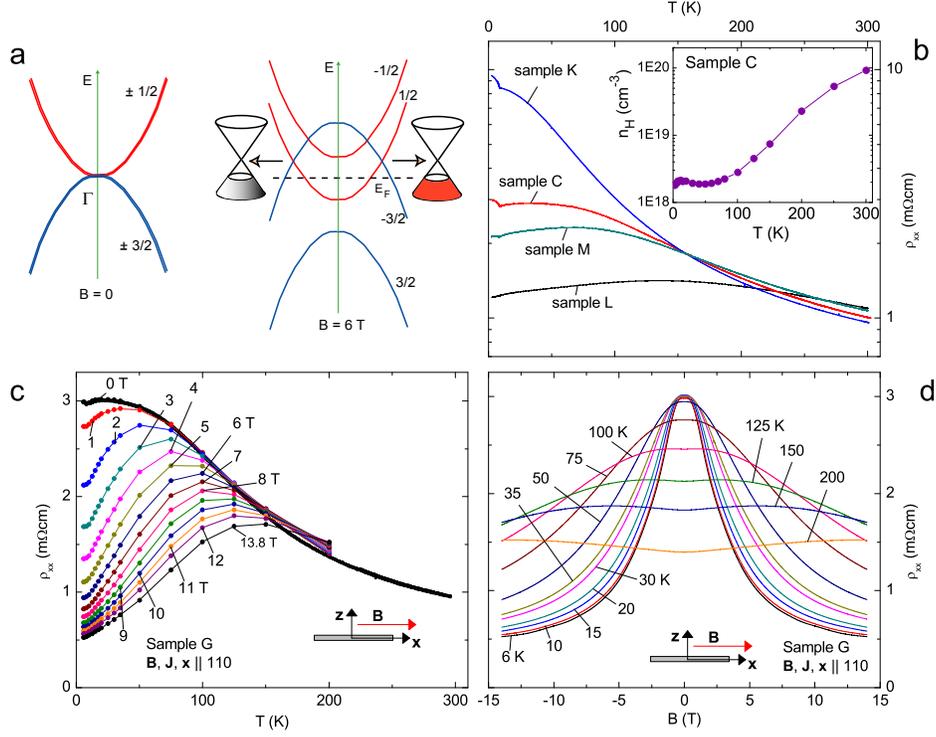}
\caption{\label{figR} 
The field-induced band crossing in GdBiPt and the chiral anomaly in its longitudinal magnetoresistance (LMR). 
Panel (a) (left sketch) shows the 4-fold degeneracy at $\Gamma$ in zero $B$ of the bands $|\frac32,\pm\frac32\rangle$ (blue curve) and $|\frac32,\pm\frac12\rangle$ (red). In finite $\bf B$ ($\parallel$ to axis shown in right sketch), the larger Zeeman shift of $|\frac32,\pm\frac32\rangle$ leads to Weyl nodes with opposite $\chi$ (red and grey cones).
Panel (b) plots the non-metallic resistivity profiles (at $B=0$) in Samples K, C, M and L. The Hall density $n_H$ in C falls by a factor of 50 between 300 and 2 K (inset). In a longitudinal field $\bf B\parallel J$ (see inset), $\rho_{xx}$ decreases with increasing $B$ below 170 K (Panel (c)). In Panel (d), $\rho_{xx}(B)$ shows a negative LMR at 6 K with a bell-shaped profile, which remains resolvable to $>$150 K. The evidence strongly support identification of the chiral anomaly as the origin of the LMR.
}
\end{figure*}

The unit cell of GdPtBi is comprised of Pt-Gd tetrahedra arrayed in the zincblende structure (Supplementary Information Sec. S1). The low-lying states involve only the four Bi $6p$ bands, $|j, m_j \rangle = |\frac32,\pm\frac12\rangle$ and $|\frac32, \pm \frac32 \rangle$, which are 4-fold degenerate at the $\Gamma$ point (the lattice has $T_d$ symmetry at $\Gamma$). Combining \emph{ab initio} calculations (Methods) with the $\bf k\cdot p$ model in finite $\bf B$, we find that the Zeeman energy results in crossings (Fig. \ref{figR}a). The number of low-lying nodes depends on whether ${\bf B}$ is aligned $\parallel$ [110] or [111] (Supplemental Fig. S4)). Details of the $\bf k\cdot p$ calculations are given in the Supplemental Sec. S2 and in Ref.~\cite{Cano}.

\begin{figure*}[t]
\includegraphics[width=16 cm]{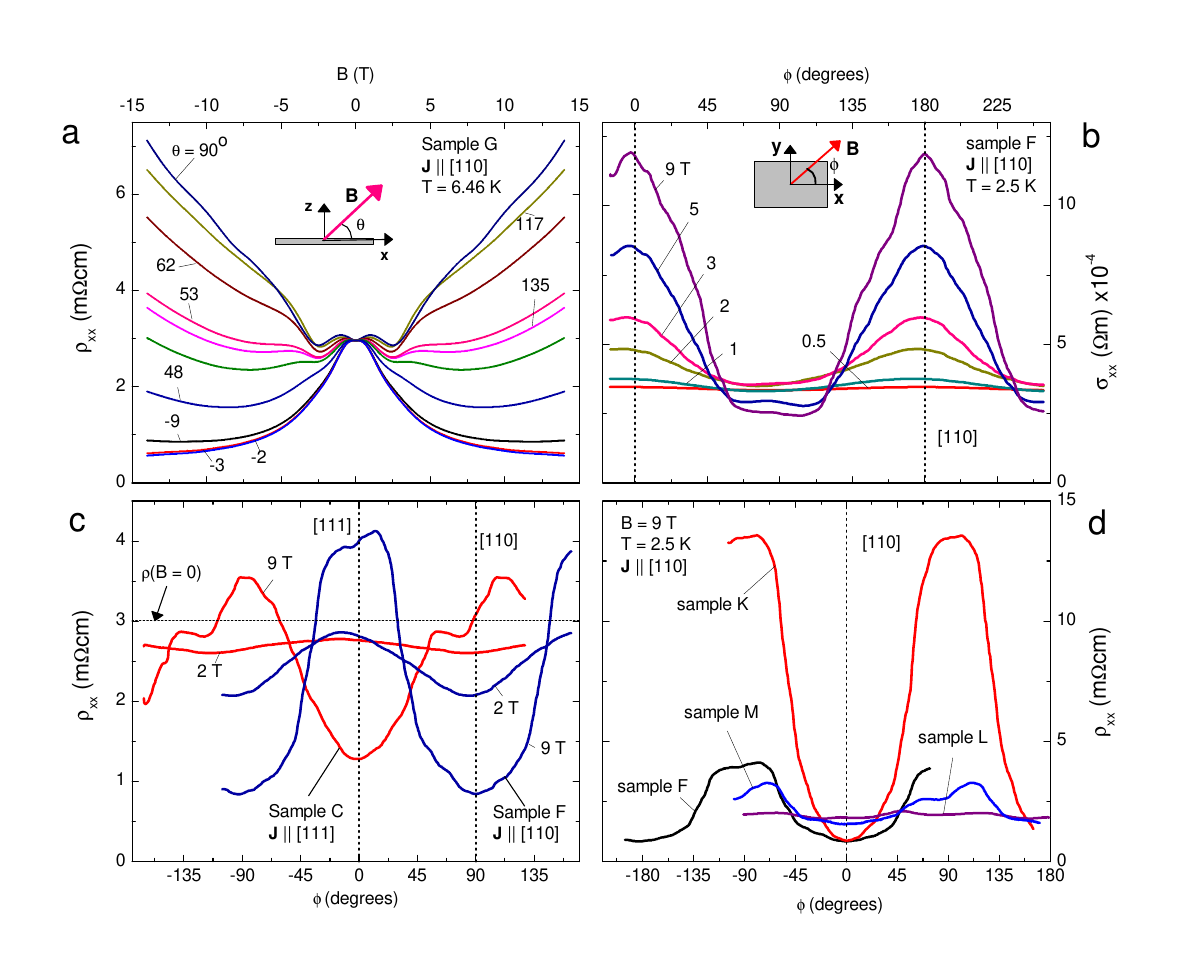}
\caption{\label{figRot} 
The dependence of the LMR in GdBiPt on field-tilt angles ($\theta, \phi$). Panel (a) shows curves of $\rho_{xx}$ vs. $B$ in G at 6.46 K, for selected values of $\theta$ (angle between $\bf B$ and $\bf J$ in polar plane $x$-$z$; see inset). As $\theta$ deviates from 0, the bell-shaped profile broadens. For $\theta\sim 90^\circ$, the MR is positive. The low-field oscillations reflect changes caused by Zeeman shift of the bands. Panel (b) shows the angular width of the conductivity $\sigma_{xx}$ vs. azimuthal angle $\phi$ for selected $B$ in Sample F at 2. 5 K ($\phi$ is defined in inset). The ``axial plume'' is similar to that in Na3Bi, but broader in angular width. Panel (c) illustrates field steering by comparing the angular plots in Sample C (with ${\bf\hat{x}\parallel }[111])$ and F ($\parallel[110]$) at two values of $B$ (2 and 9 T). In both cases, the LMR is seen only when $\bf B$ aligns with $\bf J\parallel \hat{x}$. Panel (d) shows that the magnitude of the negative LMR is large in K ($E_F$ closest to the node), but steadily decreases as $|E_F|$ moves away from the node (Samples F, M and L), consistent with the chiral anomaly.
}
\end{figure*}

Crystals of GdBiPt were cut with the axis $\bf\hat{x}$ of the longest edge either $\parallel$ [110] or $\parallel$ [111] (see Methods). Altogether, we measured 15 samples (A, B, $\cdots$, Q), with the current density $\bf J$ or heat current density ${\bf J}_Q$ applied $\parallel {\bf\hat{x}}$ in all samples (see Supplemental Table 1). Samples C, E, F and G (cut from the same boule) have very similar carrier densities. Figure \ref{figR}b shows curves of $\rho$ vs. $T$ in zero $B$ in samples K, C and M with Fermi energy $E_F<0$ ($p$-type) and one (L) that is $n$-type. The non-metallic profile and the sharp decrease of the Hall density $n_H$ are consistent with a zero-gap material (Fig. \ref{figR}b, inset). At 4 K, all samples display a prominent negative LMR measured with $\bf B\parallel E\parallel \hat{x}$. Figure \ref{figR}c plots the longitudinal resistivity $\rho_{xx}$ vs. $T$ at selected values of ${\bf B}\parallel [110]$ in Sample G. The field suppression of $\rho_{xx}$ onsets at $T\sim$150 K and increases strongly as $T\to$ 2 K. The profile of $\rho_{xx}$ vs. $B$ is a bell-shaped curve with a halfwidth $\delta B\sim$ 2.5 T below 10 K (Fig. \ref{figR}d). Raising $T$ rapidly increases $\delta B$, but the negative LMR remains observable up to 150 K. The negative LMR, observable to 150 K, is unrelated to the antiferromagnetic state that appears below the N\'{e}el temperature $T_N$ = 8.8 K, which is insensitive to $n_H$ (see Supplemental Fig. S9 and Methods).

First we show that the negative LMR goes away when $\bf B$ is tilted away from $\bf E$. As shown in Fig. \ref{figRot}a, increasing the tilt angle $\theta$ rapidly broadens the bell-shaped LMR profile. As $\theta\to 90^\circ$, the MR becomes positive apart from a low-field oscillatory feature (see below). The angular variation of the plume in the conductivity $\sigma_{xx}$ at fixed $B$ (Fig. \ref{figRot}b) is consistent with the chiral anomaly (the plume here is slightly broader than that observed in Na$_3$Bi~\cite{Xiong}). In weak $B$, the LMR profile differs between $\bf B\parallel$[111] and $[1\bar{1}1]$ consistent with band calculations (see Supplemental Sec. S2 and Figs. S2 and S13). 

A striking property of the anomaly is that when the direction of $\bf E\parallel J$ is rotated to a new crystal axis, the plume direction moves to the new axis. In Ref. \cite{Xiong}, $\bf E$ was changed by selecting a different pair of current contacts on the same crystal. Here, we compare $\rho_{xx}$ measured in the 2 crystals measured with $\bf J$ aligned with [110] vs [111]. Figure \ref{figRot}c plots $\rho_{xx}$ versus $\phi$ (the angle between $\bf B$ and $\bf J$ in both cases) for $B$ fixed at 2 and 9 T. In both crystals, $\rho_{xx}$ attains a minimum only when $\bf B\parallel J$, consistent with the chiral anomaly. (We have also confirmed (Supplemental Fig. S11) the field-steering property using just one crystal and alternating the current contacts as in Ref. \cite{Xiong}.) The results together confirm the field-steering property.

\begin{figure*}[t]
\includegraphics[width=16 cm]{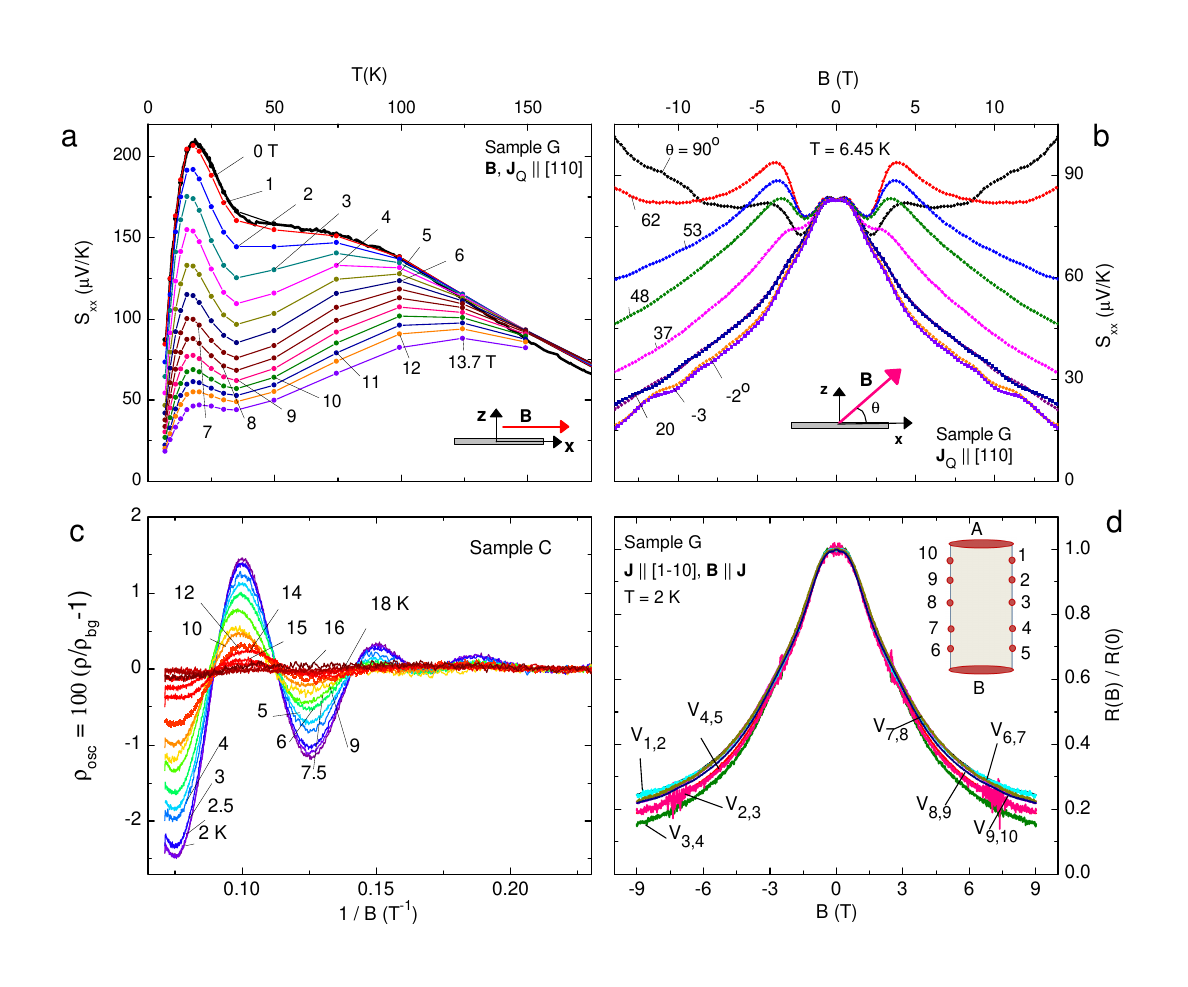}
\caption{\label{figS} 
The variation of the thermopower with $B$ and quantum oscillations in GdBiPt. Panel (a) plots the Seebeck coefficient $S_{xx}(B)$ vs. $T$ in G at selected $B$ with ${\bf B}\parallel \hat{x}\parallel{\bf J}_Q$ (the heat current density). The field suppression of $S_{xx}$, largest at 6 K is resolvable up to 150 K. In zero-$B$, the $T$-linear dependence of $S\equiv S_{xx}(0)$ below 15 K yields $E_F$ = 3.1 meV. Panel (b) shows how the field profile of $S_{xx}$ at 6.45 K changes when $\bf B$ is tilted by angle $\theta$ to $\bf\hat{x}$ in the $x$-$z$ plane (see inset). In the curves at $\theta = -2^\circ, -3^\circ$ quantum oscillations appear above 6 T. In Panel (c), plots of the oscillatory part of $\rho_{xx}$ (expressed as a percent) in Sample C show the damping of the SdH oscillations with increasing $T$. Panel (d) plots the resistances $R(B)$ inferred from the potential drops $V_{i,j}$ across 8 pairs of nearest-neighbor pairs (inset). The close similarity of the LMR profiles is evidence that the LMR is intrinsic, rather caused by distortions of current paths in a disordered crystal.
}
\end{figure*}

As a third check, we verify that the LMR is suppressed when $E_F$ is far from the node. When $|E_F|$ exceeds the Weyl energy scale, the chiral anomaly becomes unresolvable. Using the zero-$B$ thermopower $S$, $n_H$ and the SdH period, we have determined $E_F$ in 6 crystals. The magnitude of the LMR, measured by $\rho(0 T)/\rho(9 T)$, is largest when $|E_F|$ is closest to the node (sample K with LMR = 11). This variation is plotted in Supplemental Fig. S16 (also Figs. S14 and S15). These tests confirm that the LMR is associated with the chiral anomaly (they also establish that the LMR is distinct from the isotropic MR seen in gapped half-Heuslers~\cite{Casper}).

Lastly, we show that the LMR is an intrinsic effect rather than a spurious consequence of inhomogeneous ${\bf J(x)}$ caused by disorder (Methods). In Sample G, we retained the large current contacts (A and B) and added small voltage contacts (1,$\dots$,10) (Fig. \ref{figS}d, inset and Supplemental Fig. S18). As plotted in Fig. \ref{figS}d, the potential drops $V_{i,j}$ (where ${i,j}$ run over the 8 nearest-neighbor pairs) all show closely similar LMR profiles. Furthermore, from extensive simulations (Supplemental Sec. S7 and Fig. S21), we find that the mobility $\mu$ estimated from the Hall angle $\theta_H$ is far too small for ``current jetting'' to be the origin of the LMR (from Supplemental Fig. S12a, $\mu$ = 1,500 cm$^2$/Vs at 6 K).

GdBiPt provides a platform to explore the thermoelectric response of Weyl fermions. Figure \ref{figS}a shows the $T$ dependence of the Seebeck coefficient $S_{xx}$ measured in sample G with $\bf B\parallel \hat{x}$ (in all samples, both the thermal gradient $-\nabla T$ and ${\bf J}_Q$ are $\parallel \hat{\bf x}$). In longitudinal $\bf B$, $S_{xx}$ is strongly suppressed starting at 150 K high above $T_N$. The suppression of $S_{xx}$ is highly directional, rapidly diminishing as $\bf B$ is tilted away from $\bf\hat{x}$ (Fig. \ref{figS}b).

\begin{figure*}[t]
\includegraphics[width=16 cm]{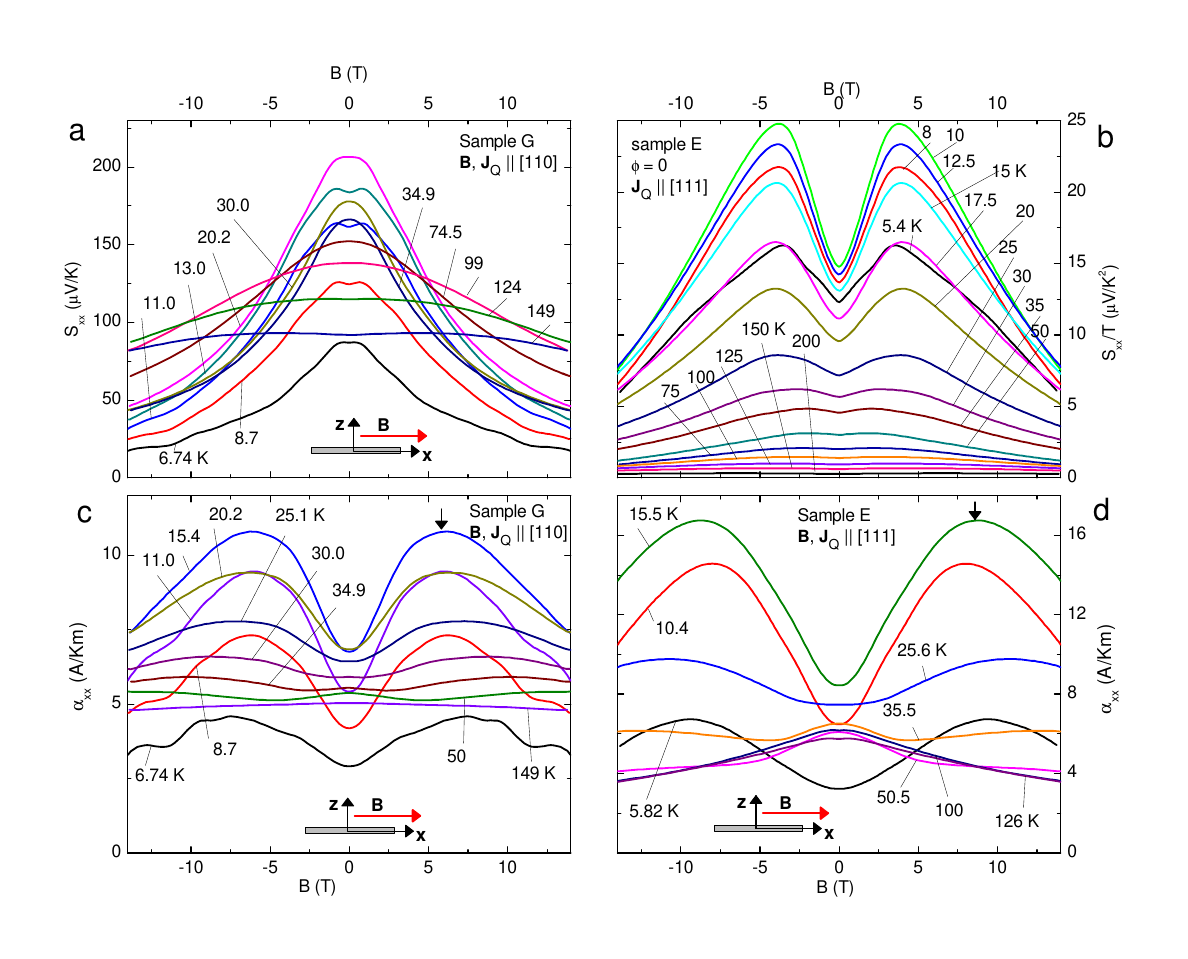}
\caption{\label{figA} 
Anisotropy of the thermoelectric response in GdBiPt in longitudinal field. Panel (a) plots the curves of $S_{xx}$ vs. $B$ in Sample G (with ${\bf B \parallel J}_Q\parallel [110])$. The curves decrease monotonically with $B$ at all $T$ (apart from a tiny dimple in zero $B$ below 20 K). Quantum oscillations are resolvable in the curve at 6.74 K. By contrast, the curves of $S_{xx}(B)/T$ in Sample E in Panel (b) ($\bf B\parallel [111])$ are non-monotonic (we plot $S_{xx}/T$ to minimize overlap). Below 25 K, the low-field region in E is dominated by a $V$-shaped profile. Unlike $S_{xx}(B)$, the curves of the thermoelectric response function $\alpha_{xx} = S_{xx}/\rho_{xx}$ are broadly similar below 30 K between the 2 geometries (Panels (c) and (d)). In Panel (c) $\alpha_{xx}$ in Sample G rises to a broad maximum at $B_p\simeq$ 6 T (arrow) before falling steeply. Quantum oscillations appear for $B>B_p$. In Panel (d) $\alpha_{xx}$ shows a similar profile except that $B_p\simeq$ 8 T. 
}
\end{figure*}

We first focus on the zero-$B$ curve in Fig. \ref{figS}a. As $T$ decreases from 300 K, the (hole-like) thermopower $S$ rises monotonically, accelerating below 40 K to attain a large peak value of 215 $\mu$V/K at 18 K. Below 15 K, $S$ decreases linearly with $T$, consistent with a metal with an unusually small $E_F$. 
Using the Mott relation $S = (\pi^2/3)(k_B^2/e) (\beta T/E_F)$, we infer $E_F$ = 3.1 meV ($\beta\sim$1.5 an exponent). Combining this with the weak-$B$ Hall coefficient $R_H$, we calculate an effective mass $m^*/m_0 \simeq 1.8$. In large field ($>$6 T), the onset of SdH oscillations provides an independent determination of the cyclotron mass $m_{cyc}$ from the damping of the SdH amplitudes with $T$ (Figs. \ref{figS}c,d and Supplemental Fig. S12b). Interestingly, we find that $m_{cyc}/m_0 = 0.23\pm 0.03$ (8$\times$ smaller than $m^*$). This large discrepancy -- unexpected in a conventional metal --  is strong evidence that the moderately heavy mass in zero $B$ changes to the small mass of Dirac states, as predicted in the induced Weyl node picture.

As noted in Fig. \ref{figS}b, the field suppression of $S_{xx}$ is strongest when $\bf B$ is aligned with $-\nabla T$ ($\theta\to 0$). The half-widths of the bell-shaped profiles $\delta B$ broaden rapidly as $T$ is raised above 80 K. Also, the suppression of $S_{xx}$ goes away in a transverse field. These features imply that the strong suppression of $S_{xx}$ arises from the chiral anomaly. 

At low $B$, however, the process of node formation (which leads to a sharp peak in the density of states ${\cal N}(E)$ near $E_F$; Supplemental Fig. S5) differs between the two cases ${\bf B}\parallel [110]$ versus [111]. The differences may underlie the strong anisotropy observed in $S_{xx}$. Figures \ref{figA}a and \ref{figA}b compare the field profiles of $S_{xx}(B)$ in Sample G (${\bf B}\parallel [110]$) and in E (${\bf B}\parallel [111]$). In Sample G, $S_{xx}$ decreases monotonically with $B$ at all $T$ (apart from a tiny dimple in weak $B$). By contrast, in Sample E, $S_{xx}$ initially increases in weak $B$, attaining peaks at $\pm$3.5 T before decreasing rapidly in large $B$. 
 
This striking anisotropy may reflect the differences in the thermoelectric response function $\alpha_{xx}(B)$ which directly relates the gradient to $\bf J$ via $J_i = \alpha_{ij}(-\partial_j T)$ ($\alpha_{xx} = S_{xx}(B)/\rho_{xx}(B)$). For Weyl fermions, recent Boltzmann equation calculations yield for $\alpha_{xx}$ (at one node) ~\cite{Fiete,Tewari,Spivak}
\be
\alpha_{xx} = e\tau\int\frac{d^3k}{(2\pi)^3} \frac{\partial f_0}{\partial E_{\bf k}} \frac{(E_{\bf k}-E_F)}{T}
\frac{[ {\bf v_k} + e{\bf B\, v_k\cdot} \mathbf{\Omega}]^2}{(1+ e{\bf B\cdot}\mathbf{\Omega})},
\label{alpha}
\ee 
where $f_0$ is the Fermi-Dirac distribution, $E_{\bf k}$ the energy, and $\bf v_k$ is the band velocity in state $\bf k$ with $\tau$ the relaxation time.

In Figs. \ref{figA}a and b, we compare the profiles of $\alpha_{xx}$ vs. $B$ in Sample G measured with ${\bf B}\parallel-\nabla T\parallel[110]$ against Sample E (${\bf B}\parallel-\nabla T\parallel [111])$. Below 30 K, the profiles in both samples are qualitatively similar: $\alpha_{xx}(B)$ increases as $B^2$ to attain a broad maximum at the peak field $B_p$, which separates two distinct field regimes.

In low fields ($B<B_p$), the changes to ${\cal N}(E)$ lead to an increasing $\alpha_{xx}$ in both geometries. The difference in $B_p$ (6 vs. 8 T) suffices to produce the $V$-shaped profile in $S_{xx}(B)$ of Sample E, but not in G. The increase in $\alpha_{xx}$ in the range $0<B<B_p$ is a signature of the node creation process that is not understood.

In the regime above $B_p$ the Weyl nodes are fully formed with well-resolved Landau levels (LLs) as indicated by the quantum oscillations (curves below 11 K in Fig. \ref{figA}a). The dominant feature is the steady decrease in $\alpha_{xx}$ with increasing $B$ which drives $S_{xx}$ towards zero at large $B$ (in G, we enter the $n$ = 0 LL above 25 T). A characteristic of Weyl states in the quantum limit is the one-dimensional dispersion along the $\bf B$ axis, which implies a density of states ${\cal N}\sim eB/v$ that is $E$ independent. Consequently, $\alpha_{xx}\sim \partial {\cal N}/\partial E$ vanishes (by contrast $S_{xx}$ increases linearly with $B$ in the lowest LL for a massive Dirac system~\cite{Liang}). We interpret the downward trend in both $\alpha_{xx}$ and $S_{xx}$ above $B_p$ as consistent with charge pumping between the Weyl nodes associated with the chiral anomaly. The curves of $\alpha_{xx}$ can provide sharp tests of Eq. \ref{alpha} in the intermediate field regime.

We have observed in GdBiPt a large, negative longitudinal MR when $E_F$ is near zero. By varying both the directions of $\bf B$ and $\bf E$, we confirm that the enhanced conductance is confined to a plume centered at $\bf B\parallel E$. Moreover, it is steerable by the direction of $\bf B$. Finally, we show that the LMR is most prominent when $|E_F|$ is closest to the node. The 3 tests together with the measurements of $m^*$ present a strong case for the chiral anomaly in Weyl nodes. The observed thermoelectric response function $\alpha_{xx}$ shows a decrease in large $B$, consistent with ${\cal N}$ of a chiral $n=0$ LL. The anisotropy of the thermoelectric response is consistent with the anisotropic nature of the node creation, although a full accounting of the results awaits further analysis. More broadly, we have shown that field-induced crossing of degenerate bands can result in Weyl nodes and the associated chiral anomaly. The results imply that zero-gap semiconductors with large spin-orbit interaction (e.g. half-Heuslers, gray tin and HgCdTe) are prime candidates for exploring Weyl physics.

\newpage



\vspace{5mm}\noindent
{\bf Supplementary Information} is available in the online version of the paper.

\vspace{5mm}\noindent
{\bf Acknowledgements}  
We are indebted to Jennifer Cano, Barry Bradlyn and Jun Xiong for discussions, and Seyed Koohpayeh, Jason Krizan, and Weiwei Xie for technical assistance. The research is supported by a MURI award for topological insulators (ARO W911NF-12-1-0461) and by the Army Research Office (ARO W911NF-11-1-0379). The growth and characterization of crystals were performed by S.K and R.J.C., with support from the National Science Foundation (NSF MRSEC grant DMR 1420541). C.A.B. was an REU participant funded by the NSF-MRSEC grant DMR 1420541. N.P.O. acknowledges the support of the Gordon and Betty Moore Foundation’s EPiQS Initiative through Grant GBMF4539. B.A.B. acknowledges support by NSF CAREER DMR-095242, ONR-N00014-11-1-0635, NSF grant DMR 1420541, Packard Foundation and Keck grant.

\vspace{5mm}\noindent
{\bf Author Contributions} 
M.H. performed most of the measurements with early assistance from C.A.B. The crystals were grown and characterized by S.K. and R.J.C. Analyses of the results were done by M.H., Z.J.W., Q.G., B.A.B. and N.P.O. Simulations of current-distributions were performed by S.H.L. The manuscript was written by M.H. and N.P.O. with contributions from all authors.

\vspace{5mm}\noindent
{\bf Author Information} The authors declare no competing financial interests. Correspondence and requests for materials should be addressed to M.H. (hirschberger.max@gmail.com) and N.P.O. (npo@princeton.edu).

{\ddag} Address of C.A.B.: Wellesley College, 106 Central Street, Wellesley, MA 02481


\vspace{10mm}
\noindent {\bf  Methods}\\
Electronic structure calculations were performed in the framework of density functional theory (DFT) using the WIEN2K code~\cite{wien2k}. We used a full-potential linearized augmented plane-wave and local orbitals basis with the PBE parametrization of the generalized gradient approximation~\cite{Perdew1996}. The plane-wave cutoff parameter $R_{MT}$ $K_{max}$ was set to 7. The modified Becke Johnson (mBJ) functional~\cite{Becke2006} was used. Spin orbit coupling (SOC) was included in the calculation as a perturbative step. A large exchange parameter $U_\text{eff} = 7\,$eV was applied to the Gd $4$f states (with “open core” treatment of the 4$f$ electrons), in effect removing these states far away from the Fermi energy $E_F$.

Single crystals of pure and Au-doped GdPtBi were grown using Bi-self flux. Stoichiometric mixtures of the elements were placed in an alumina crucible and sealed inside a quartz tube under vacuum. The ampoules were heated to $1130\,^\circ$C and kept for $12$ hours at constant temperature to get a homogeneous solution. The samples were cooled at the rate of $1.5\,^\circ$C/hr, to $940\,^\circ$ C and then centrifuged to remove Bi-flux. Single crystals of ~2 mm in size were successfully obtained. 

We have used the Hall effect to characterize our crystals, and to obtain estimates of the carrier density, assuming conduction from a single band at low temperature ($\sim$4 K). The Hall resistivity at 4 K was fitted in the low-field limit to a line to extract the Hall density $n_H = 1/(e R_H)$, where $R_H$ the Hall coefficient. At high temperatures, thermally activated carriers suppress the amplitude of $R_H$ (enhance $n_H$), but the saturated value of $n_H$ at low $T$ may be used to estimate the intrinsic carrier concentration (results are shown in Supplement). The onset of the low-temperature regime where $n_H$ saturates is expected to scale with the Fermi energy $E_F$ in a zero-gap material.

Altogether, we investigated single crystals from three batches of pristine GdPtBi (batches 1-3), all of which exhibited intrinsic $p$-type behavior in the Hall effect and in the thermopower. The highest quality crystals (estimated by the amplitude of quantum oscillations, and the enhancement of the thermal conductivity $\kappa_{xx}$ at low temperature) have the lowest carrier concentrations. We also investigated 3 batches of crystals doped slightly with Au ($0.5$, $5$, $10\,\%$ for batches 4, 5, 6 respectively) to substitute for Pt. We found that only a small fraction of the Au in starting materials is incorporated in the final crystals, as determined by measurements of the Hall density. For example, the low-temperature Hall densities measured for the samples in batch 5 are (1.4 --3.5)$\times 10^{18}\,1/$cm$^3$. But if $5\,\%$ of the Pt had been replaced by Au, we would expect densities of around $-6.7 \times 10^{20}/$cm$^3$, using the experimental lattice parameter $a = 6.68\,$\AA~\cite{}. Batch 4 contained crystals which had very low Hall densities (close to compensation).

The so-called ``18 electron rule'' for half-Heusler materials~\cite{Dwight1974} implies that heavy doping with atoms of a different valence count is prohibited. In practice, this implies a soluability limit for Au doping in our case. We have found that crystals from batch 5 ($5\,\%$ Au for Pt in the starting boule) have similar Hall densities to batch 6 ($10\,\%$ Au), i.e. $-n_H = 1.5\cdot 10^{18}$ to $4\cdot 10^{18}\,$cm$^{-3}$.

The relatively large sizes of the single crystals used for thermopower experiments ($\sim 2\,$mm on a side) allowed us to cut and orient samples accurately using the Laue x-ray setup at Johns Hopkins University, Baltimore, MD (with the assistance of S. M. Koohpayeh and J. W. Krizan). In particular, samples C, D, E, F, G were oriented in this way.

For all other crystals, we used the edges of hexagonal facets of the single crystals ([111] direction of the cubic structure) to determine the orientation; the angular error of cut planes may be larger for such samples, and we estimate it at $\delta \phi \sim 10\,$degrees. These samples may be as small as $\sim 0.5\,$mm. After cutting and polishing, electrical contacts to the platelets (thickness $t\sim 0.1\,$mm) were made by silver paint and thin gold wires. In some cases, the contacts cover a considerable fraction of the surface area. This implies limitations of determining the correct sample geometry (using an optical microscope) for calculating the Hall resistivity $\rho_{yx}$, and by extension the Hall density $n_H$. An error of up to $\delta n_H/n_H\sim 30\,$\% may be incurred for the smallest crystals.

To test for inhomogeneous current distribution, we remounted Sample G, replacing the previous voltage contacts with 10 new, small voltage contact pads. The current contacts A and B are sufficiently large to cover the shorter edges of the crystal. At $T$ = 2 K, we measured simultaneously the potential difference $V_{i-j}$ between the 8 pairs of nearest-neighbor contacts ($V_{1-2}$, $V_{2-3}, \cdots, V_{9-10}$) as $B$ (applied $\parallel {\bf J}\parallel {\bf \hat{x}}$) is varied.

The 8 curves for the relative change in $V_{i-j}$ are nearly identical below 3 T, only displaying slight, non-systematic deviations above 5 T (Fig. \ref{figS}d and Supplement). To us, the striking agreement across the 8 contacts provides strong evidence for the uniformity of $\bf J(r)$ throughout the crystal. From the 8 curves, we infer that the current density is uniform throughout the crystal in the LMR experiment. This implies that the observed negative LMR is an intrinsic electronic effect rather than arising from strong distortions of the current paths.

A final concern is ``current jetting'', which can lead to field-induced changes in the longitudinal MR in very high mobility semimetals and metals. 
To see if current jetting can be a plausible origin of the negative LMR observed in GdPtBi, we have performed extensive simulations for our sample geometry and mobility values $\mu$ = 1,500 -- 2,000 cm$^2$/Vs. From the equations $\nabla \cdot {\bf J} = 0$ and $J_i = \sigma_{ij}E_j$, where $E_j = -\partial_j\psi$, the potential function $\psi(x,y)$ satisfies the anisotropic 2D Laplace equation (see Supplement). The simulations show that, below 10 T, current jetting has a negligible effect (a few $\%$) in a sample with $\mu$ = 2,000 cm$^2$/Vs. This is far too small to account for the large LMR in GdPtBi. Note that from Fig. \ref{figS}d, $\rho_{xx}$ has decreased by 2$\times$ at the low field of 2 T. We would need $B>$50 T to achieve the same suppression if current jetting were the origin (see simulations in Supplement).\\

For references see Supplementary Information.

\newpage
\noindent\vspace{10mm}

{\bf {\Large Supplementary Information:}\\
The chiral anomaly and thermopower of Weyl fermions in the half-Heusler GdPtBi 
}

\section{ab initio band calculations}
\label{sec:dft}

\begin{figure}[htb]
  \centering
  \includegraphics[width=7.5cm]{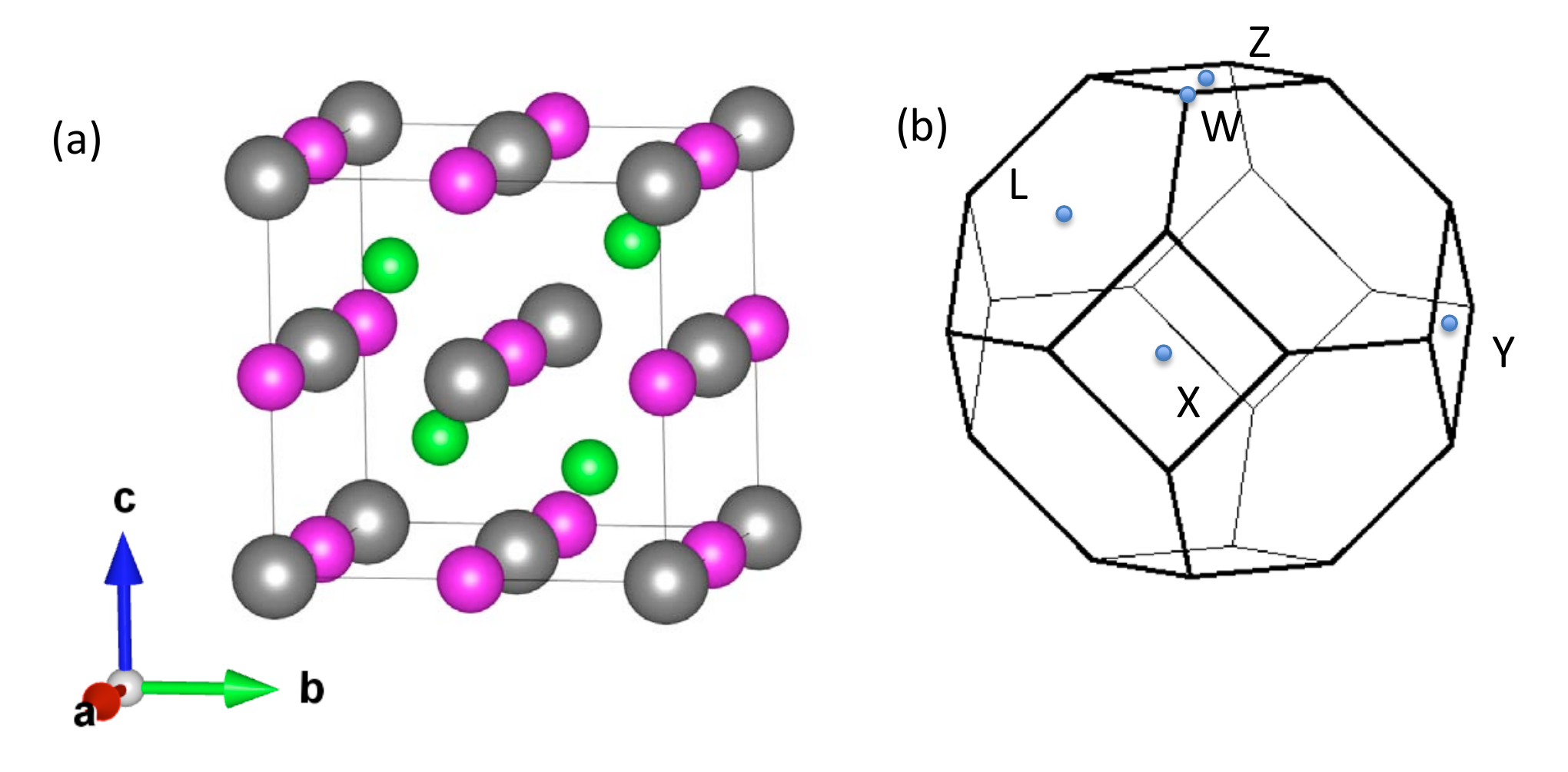}
  \caption{(a) The cubic cell of GdPtBi, containing four formula units / primitive cells. Gray, red and green balls indicate Gd, Bi and Pt atoms, respectively. (b) The first Brillouin zone of the FCC lattice.}
\label{fig:dftcrys}
\end{figure}

The crystal lattice of GdPtBi and its first Brillouin Zone are displayed in Fig. \ref{fig:dftcrys}.

\begin{figure}[htb]
  \centering
  \includegraphics[trim={3cm 15cm 9cm 0},width=4.4cm]{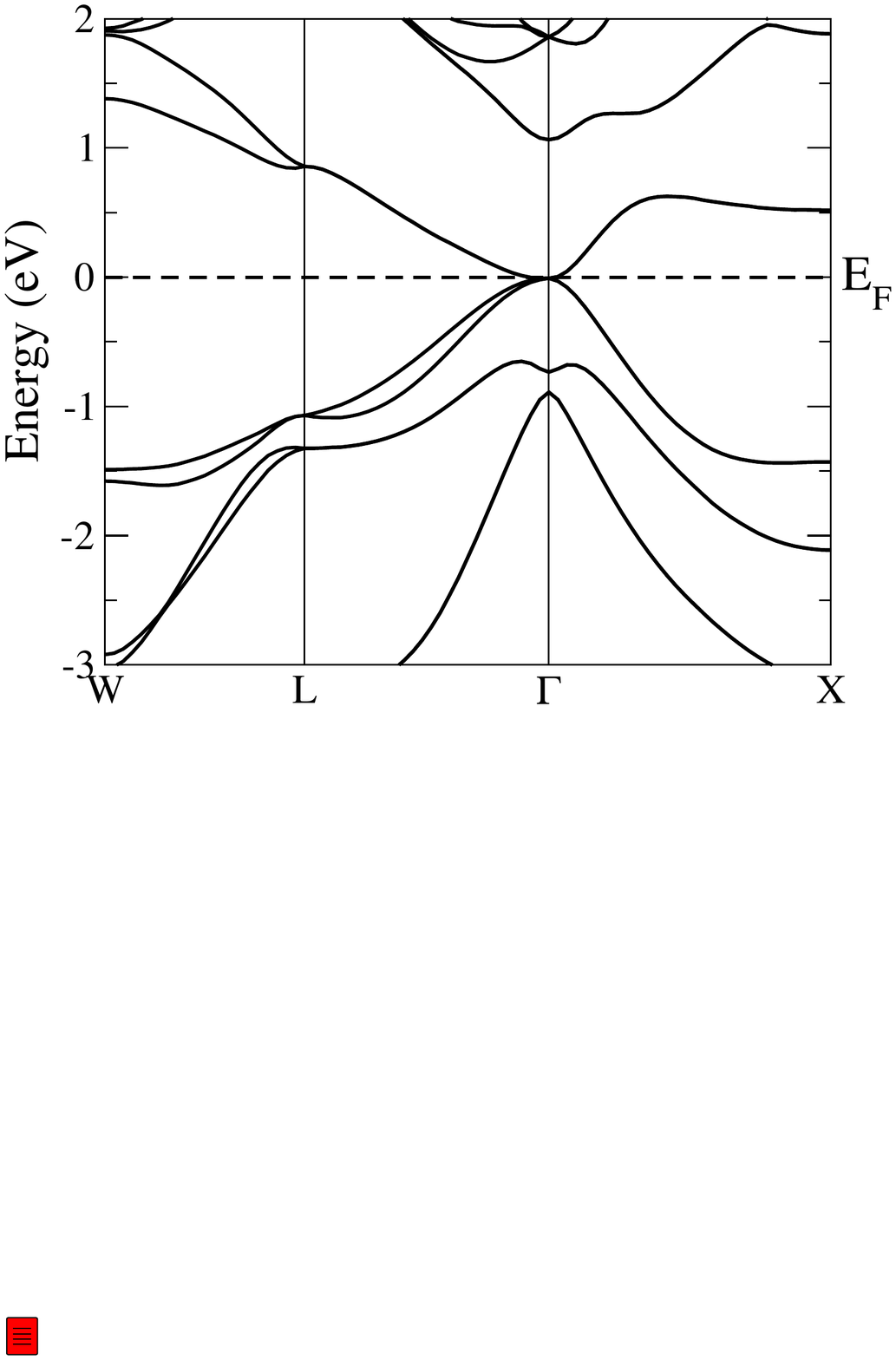}
  \caption{Calculated band structure from density functional theory (zero magnetic field). A large $U_\text{eff}$ was used to raise the $f$-states away from the Fermi energy.}
\label{fig:dft_soc}
\end{figure}

As explained in the Methods Section, band structure calculations were performed in the framework of
density functional theory (DFT) using the WIEN2K code \cite{wien2k}. The band structure in Fig. \ref{fig:dft_soc} was calculated in the paramagnetic phase (the measurements were performed mostly at temperatures $T\gg T_N$). The $4f$ states of Gd are strongly localized and weakly hybridized with the low-lying states of interest (derived from Bi 6p and Pt 6s orbitals). Hence the dispersions of the low-lying states are insensitive to the magnetic ordering assumed for the $f$ bands (the dispersions are only weakly affected even if the $f$ states are eliminated altogether). In our LDA+U approach with $U_{eff}$ = 7 eV, the splitting of the $4f$ states into two branches straddling the Fermi energy $E_F$ leads to magnetism in the $f$ states. 
The insensitivity of the low-lying states to the $4f$ states of Gd justifies an ``open core'' treatment of the $f$ states in the calculation.

The differences between the band structures in the antiferromagnetic (AFM) and paramagnetic phases are discussed further in a forthcoming publication\cite{Cano2015}.

\section{Bands in a Zeeman field: $\textbf{k} \cdot \textbf{p}$ model}
\label{sec:kdotp}
\begin{figure}[htb]
  \centering
  \includegraphics[width=6cm]{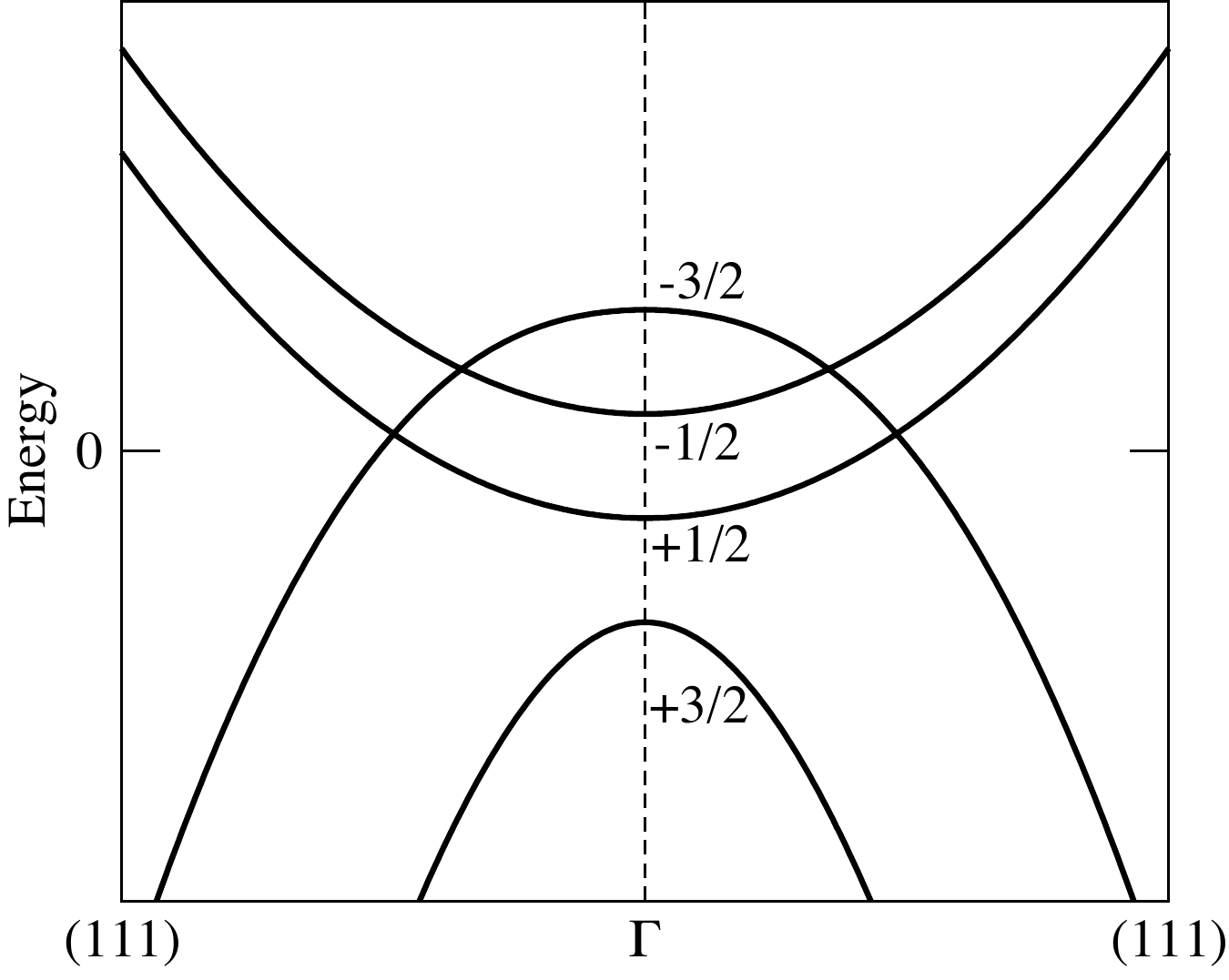}
  \caption{Schematic representation of the Zeeman split bands in finite magnetic field $\bf B$ along the [111] direction. The values of $m_j$ given here are good quantum numbers only along the direction of high symmetry.}
\label{fig:bandscross}
\end{figure}

\begin{figure}[htb]
  \centering
  \includegraphics[width=8.5cm]{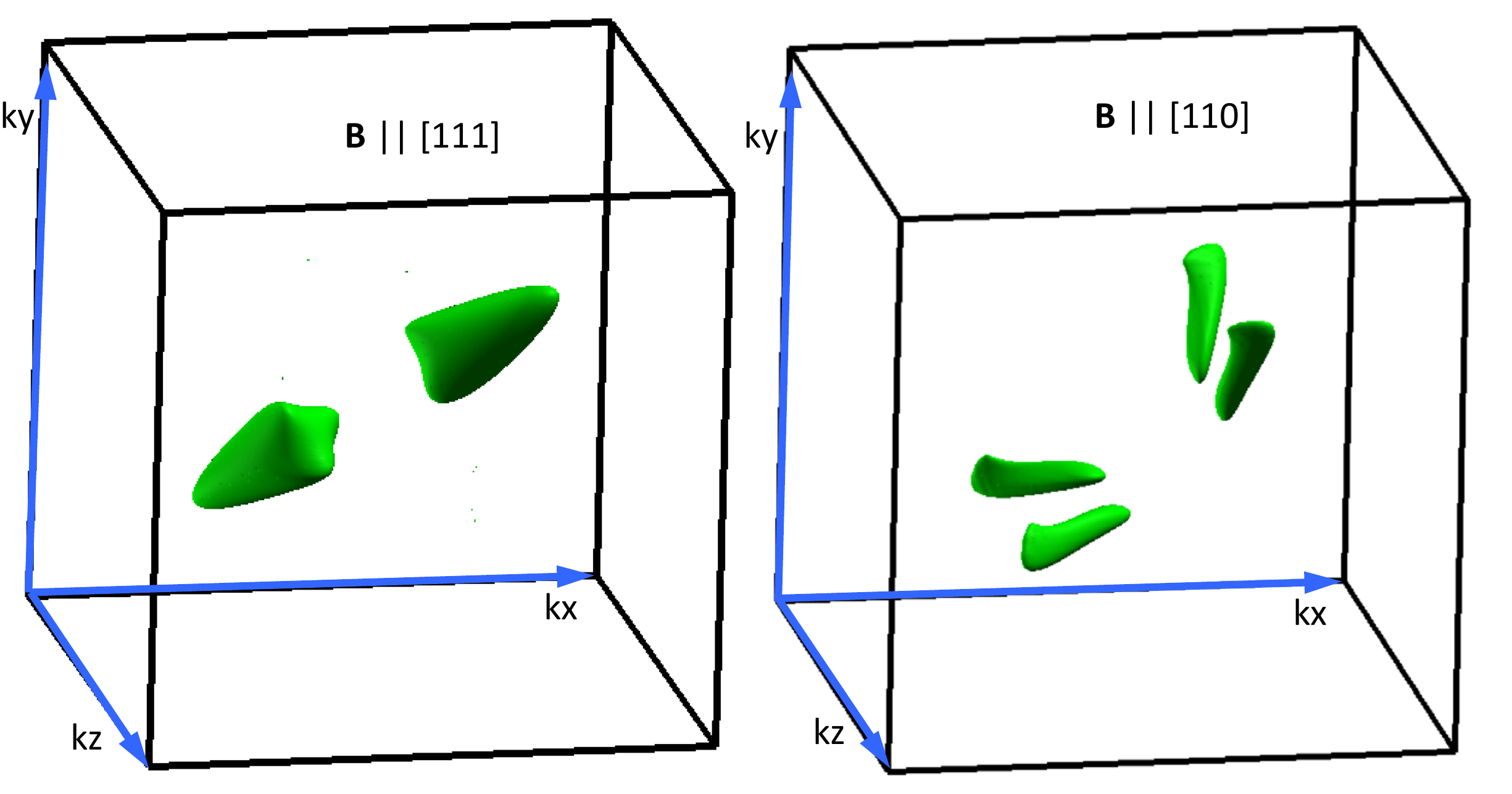}
  \caption{Calculated Fermi surface, using the $\bf k \cdot \bf p$ model, for $E_F = -10\,$meV, hypothetical $g$-factor $g=40$, and magnetic field $B = 10\,$T parallel to the [111] direction (left) and parallel to the [110] direction (right). The box displayed here extends $1/9$ of the full size of the Brillouin zone in each direction ($k_x$, $k_y$, and $k_z$) away from the $\Gamma$-point (center of the box).}
\label{fig:weylpocket}
\end{figure}

\begin{figure*}[bht]
  \centering
  \includegraphics[width=14.4cm]{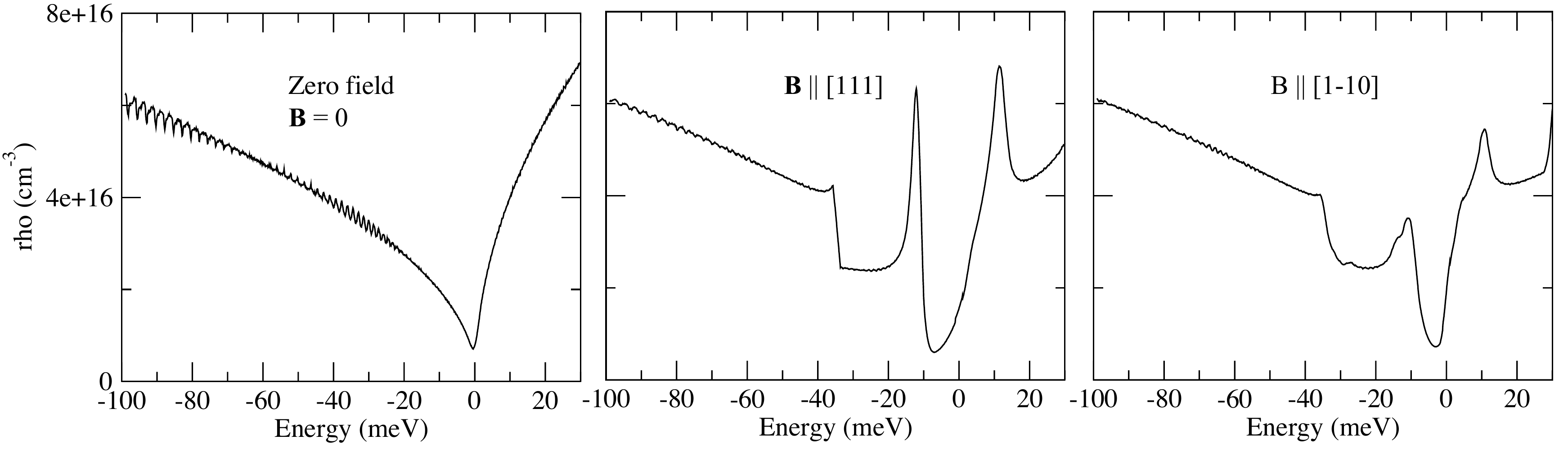}
  \caption{Calculated density of states $\rho$ as a function of energy, using the $\bf k \cdot \bf p$ model. In panels (b) and (c), we assumed a $g$-factor $g = 40$ and $B = 10\,$T parallel to the [111] direction (b) and the [110] direction (c). The sharp peak just below $E = 0$ arises due to the flat bottom of the $+1/2$-band (Fig. \ref{fig:bandscross}). The peak is more prominent for magnetic field along the [111] direction.}
\label{fig:dos}
\end{figure*}

The appearance of Weyl nodes upon application of a magnetic field $\bf B$ in GdPtBi is part of a generic phenomenon that depends only on the symmetry of the material in zero $\bf B$. While a full analysis of the problem will be presented in ref. \cite{Cano2015}, we summarize here some of the results.  GdPtBi is a material with $T_d$ symmetry group plus time-reversal ($T$) invariance. This group protects a 4-fold band crossing at the $\Gamma$ point. A symmetry-constrained $\bf{k} \cdot \bf{p}$ model valid in the vicinity of $\Gamma$ can be obtained, with parameters fitted to an \emph{ab-initio} calculation \cite{Cano2015}.  Applying a (Zeeman) magnetic field in several high-symmetry directions breaks the $T_d\otimes T$ symmetry to a subgroup and breaks the four-fold $\Gamma$ point degeneracy, as sketched in Fig. \ref{fig:bandscross}. Using the remaining  symmetries of the system, the presence of a nonzero number of Weyl points and, less generically, nodal line, can be unequivocally confirmed through either 1) band eigenvalue arguments (when the Weyl crossing is on a high-symmetry axis or when the nodal lines are on high-symmetry mirror planes),  or 2) global eigenvalue arguments that invoke Chern number differences between high-symmetry planes in the system. While the topological reasons for the existence  of Weyls can only be deduced when the field is along high-symmetry directions where the remaining symmetry group is nontrivial, the Weyls persist even when $\bf B$ is tilted away from symmetry axes. Our phase diagram \cite{Cano2015} shows that Weyl points exist \emph{for any direction} of the magnetic field $\bf B$, and for both inverted and non-inverted materials. The effect of the Weyls, whose velocity now depends on the magnetic field that created them, on the transport properties of the material are analyzed in \cite{Cano2015}.\\

We give a summary of the results for $\bf B$ applied along several high-symmetry directions (see Fig. \ref{fig:weylpocket}). When ${\bf B} \parallel [001]$ direction, two Weyl nodes appear on the high symmetry $[001]$ axis between the $J_z= -\frac12$ and $-\frac32$ bands, which have distinct $C_{2z}$ eigenvalues. On the high-symmetry $k_x=0$ plane, two Weyls are required to appear due to a more subtle Chern number argument, between  the $J_z= \frac12$ and $-\frac32$ bands. An identical number of Weyls (2) has to appear in the $k_y=0$ plane \cite{Cano2015}. This direction of the field has yet to be checked by experiment. With ${\bf B}  \parallel [111]$,  two Weyl points appear between the $\frac32$ band and the $\frac12$ band \cite{Cano2015}. Another two between the $\frac32$ band and the $-\frac12$ band, due to the eigenvalue argument. The ${\bf B} \parallel [110]$ geometry has two sets of nodal physics. Line nodes are present on a high symmetry $[1\bar{1}0]$ plane due to intersections of bands of different Mirror eigenvalues. Moreover $4$ Weyl points are found close to the $[110]$ plane (they are in the $(110)$ plane if Inversion Symmetry remains unbroken). Rotating $\bf B$ away from these high symmetry directions cannot immediately gap the stable Weyl points. Our phase diagram \cite{Cano2015} suggests the presence of nodal behavior throughout the phase space produced by arbitrary orientation of $\bf B$.

\section{Crystal growth \& structural analysis}
\label{sec:xtals}

As described in Methods, crystals of pure and Au-doped GdPtBi were grown using Bi-self flux. 
On completion of the crystal growth process, the crystals were crushed into fine powder and subjected to X-ray diffraction to obtain the powder pattern. The powder pattern is in good agreement with the reported crystal structure in the F-43m space group. The red vertical lines at the bottom of the graph correspond to the peaks of the reported cubic phase.\\

\section{Controlled Doping Experiments}
\label{sec:doping}

\begin{figure}[htb]
  \centering
  \includegraphics[width=9.2cm]{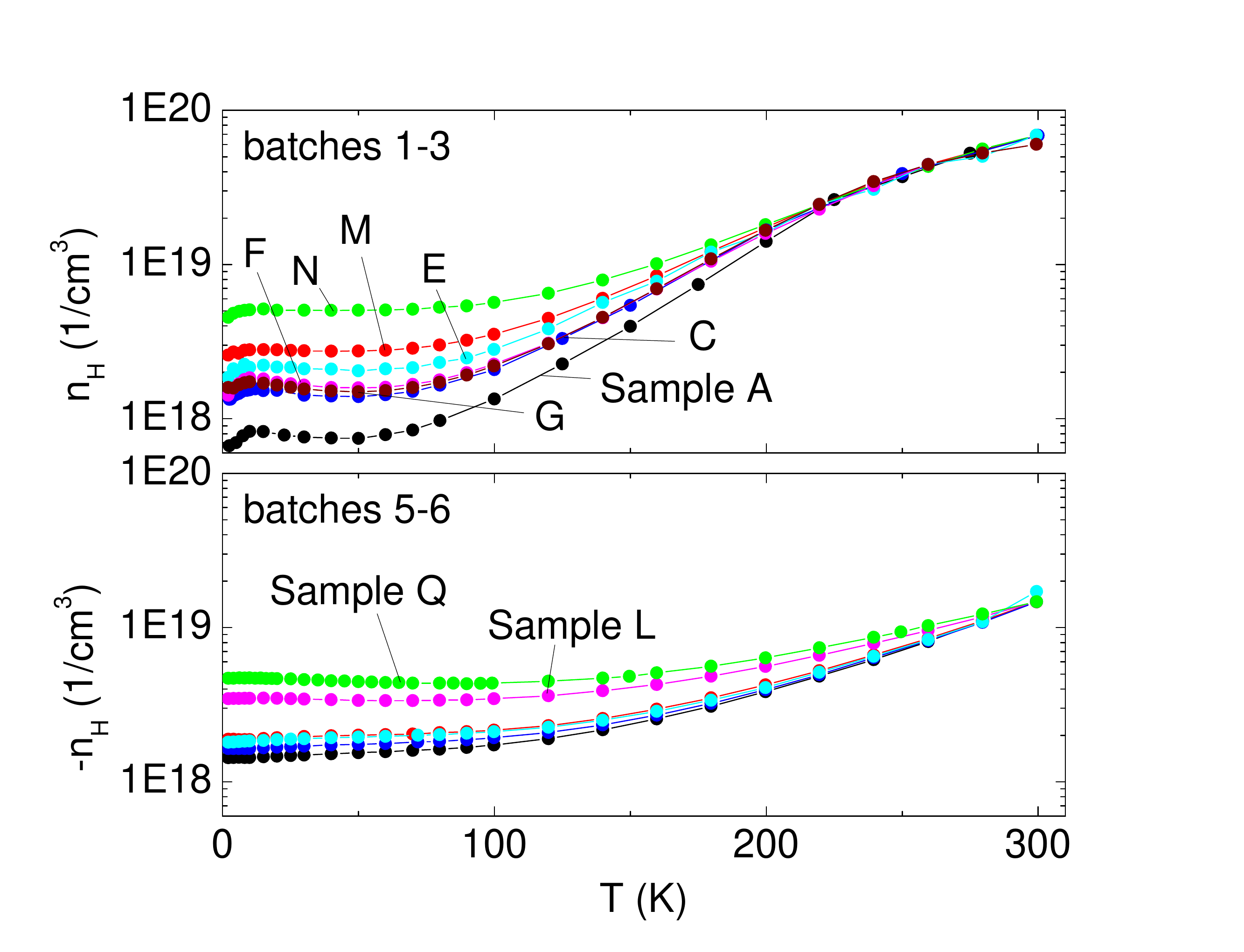}
  \caption{Hall density $n_H$ as a function of temperature for various samples. The high temperature values were scaled, to correct for errors in measuring the sample geometry (see text).}
\label{fig:hall_vtemp}
\end{figure}

\begin{table*}
	\centering
	\begin{tabular}{|c|c|c|c|c|c|c|}
	\hline
	Sample & Batch & $\bf J$-axis & $n_H (10^{18}/$cm$^3)$ & $S_F^\text{max}($T$)$ & $T_N ($K$)$ & RRR$^{-1}$ \\
	\hline\hline
	A	& 1 & $\sim 1\bar 1 0$ & $0.67$ & $28.6$	& $8.93$& $3.01$ \\
	B	&	3	& $-$ 			&	$-$ & $-$ &	$8.92$ 			& $2.45$  \\
	C	&	3	&	111				& $1.3$ & $33.1$ & $8.83$ & $2.99$   \\
	D	&	3	& $1 \bar 1 0$				& $-$ & $34.5$ & $8.71$ 	& $3.78$   \\
	E	&	3	& 111				& $1.85$ & $33.2$ & $8.83$ 	& $2.28$   \\
	F	&	3	& $1 \bar 1 0$				& $1.4$ & $33.1$ & $8.87$ 	& $3.07$   \\
	G	&	3	& $1 \bar 1 0$				& $1.8$ & $34.2$ & $8.80$ 	& $3.16$   \\
	H	&	1	& $1 \bar 1 0$				& $2.6$ & $39.9$ & $8.87$ 	& $1.56$   \\
	J	&	2	& $-$ 			& $4.5$ & $50.14$ & $8.75$ 		& $1.57$  \\		
	K	&	4	& $1 \bar 1 0$ 			& $-0.15$ & $15.22$ & $8.77$& $7.44$  \\
	L	&	5	& $1 \bar 1 0$ 			& $-3.5$ & $-$ & $8.80$ 		& $1.10$  \\			
	M	&	1	& $1 \bar 1 0$ 			& $2.6$ & $-$ & $8.82$ 			& $1.99$  \\				
	N	&	3	& $-$ 			& $-$ & $22.02$ & $-$ 			& $-$  \\						
	P	&	2	& $1 \bar 1 0$ 			& $4.5$ & $49.82$ & $8.95$ & $1.55$  \\						
	Q	&	6	& $1 \bar 1 0$ 			& $-4.7$ & $-$ & $8.88$ 				& $0.69$  \\
	\hline
	\end{tabular}
	\caption{List of named samples investigated. The sign of the Hall density $n_H$ indicates hole- or electron-like. $S_F^{max}$ is the maximum FS section inferred from the SdH oscillations. RRR is the residual resisitivity ratio. Results from some other samples, for which only limited amounts of data were obtained and which are not listed here, are also included in Figs. \ref{fig:hall_vtemp}, \ref{fig:rrrinv_vhalldens}, \ref{fig:mrrat_vhalldens}.	}
  \label{tab:samples}
\end{table*}

\begin{figure}[htb]
  \centering
  \includegraphics[width=9.2cm]{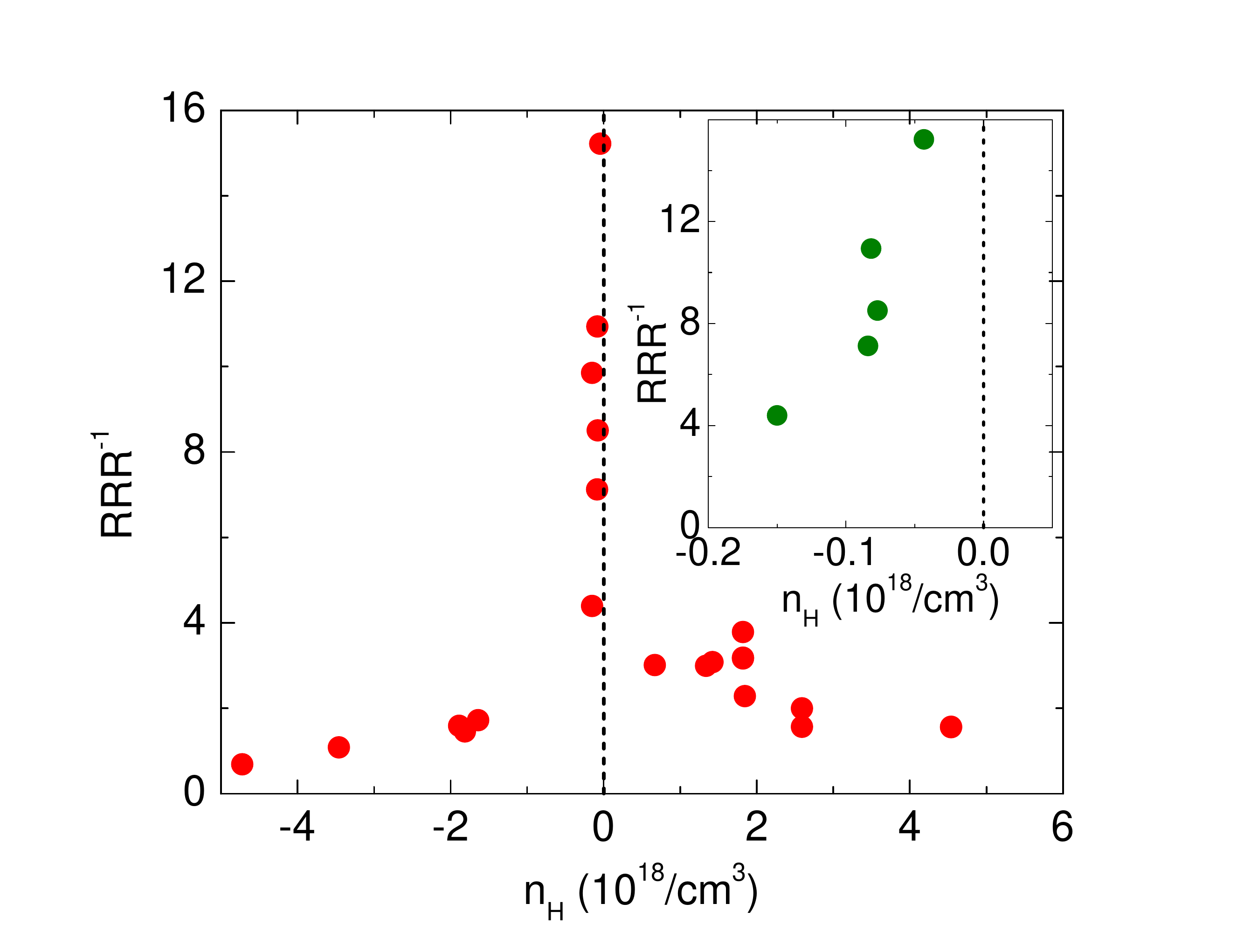}
  \caption{Inverse residual resistivity ratio (RRR) $\rho(2\,$K$)/\rho(300\,$K$)$ plotted as a function of the Hall density $n_H$ at 4 K. The most insulating samples are from batch 4. Inset: zoom-in to low negative carrier density, highlighting the samples from batch 4.}
\label{fig:rrrinv_vhalldens}
\end{figure}

Here we provide more results on the variation of Hall density $n_H$ with doping. We have scaled all the $n_H(T)$ curves for batches 1-3 to a single value at high temperatures. This value of $n_H$ (300 K) is the one obtained from the largest crystals of a particular batch, where the sample size error is expected to be the smallest. We proceeded analogously for batches 4 and 5. We believe this approach is justified by the fact that thermal broadening of the Fermi-Dirac function at high $T$ should render small variations of intrinsic carrier concentration unobservable in the transport coefficients. Consistent with the assumption, each subset of curves of $n_H(T)$ approaches the same ``universal'' shape at high $T$ (Fig. \ref{fig:hall_vtemp}).\\



We have observed quantum oscillations in most of our p-type crystals at fields $B>5$ T (with period that varies relative to the $\bf B$ axis). These results will be more thoroughly analyzed in a future publication \cite{Hirschberger2015B}. The Fermi surface area $S_F$ increases monotonically with carrier density for batches 1-3. This lends further support to our approach of scaling $n_H(T)$ at high $T$, because $S_F$ does not depend on the measurement of the sample geometry.\\

From the evolution of the $\rho(T)$ curves (RT-curves) as a function of $n_H$, we can extract two essential pieces of information: First, Fig. \ref{fig:rrrinv_vhalldens} shows that the most insulating samples are the ones with the lowest $\left| n_H \right|$, as expected for a zero-gap semiconductor.\\

Secondly, the N\'{e}el temperature $T_N$ remains unchanged (to our resolution) as $n_H$ is varied. We estimate an error of $\pm 0.1-0.2\,$K for $T_N$ as determined from the RT curves. The lack of change in $T_N$ has implications for the likely mechanism of interaction between the Gd magnetic moments. We infer that the dominant exchange driving the transition at $T_N$ only weakly affects the conduction electrons. More importantly, it shows that the changes in the thermopower and the magneto-resistance as a function of $n_H$, on which we elaborate below, cannot be a consequence of the magnetism.

To estimate the mobility below 10 K, we have used the Hall angle $\theta_H$ inferred from measurements of $\rho_{yx}$ in Sample G. Figure \ref{fig:tanHall}a plots the $T$ dependence of $\tan\theta_H/B$. Below 10 K, the carriers are predominantly hole like. Hence the mobility $\mu$ equals 1,500 cm$^2$/Vs. Above 100 K, the large population of excited $n$-type carriers adds a negative contribution which to ``cancels'' a significant part of the Hall angle signal, so a mobility value can only be inferred by resorting to a two-band fit. 

Figure \ref{fig:tanHall}b plots the damping of the SdH peak amplitudes in Sample C for 4 peak values at the fields indicated. By fitting (solid curves) to the standard Lifshitz-Kosevich expression~\cite{Roth}, we have extracted a cyclotron mass $m_c\sim 0.23 m_0$ ($m_0$ is the free electron mass). 

As anticipated by the band calculations, the process of creating the Weyl nodes by $\bf B$ is anisotropic. For ${\bf B}\parallel [111]$, two Weyl nodes are created whereas for ${\bf B}\parallel [1\bar{1}0]$, four nodes appear. Evidence for this anisotropy appears in the weak-$B$ LMR. In Fig. \ref{Zoom} we plot in expanded scale the LMR observed in 4 samples. When ${\bf B}\parallel [1\bar{1}0]$ (Samples F and G), the profile displays a nominal plateau in weak $B$ (instead of the $B^2$ behavior predicted by the Son-Spivak expression~\cite{Son}). This suggests to us that $B$ must exceed a finite threshold value before the axial current contribution becomes resolvable. For ${\bf B}\parallel [111]$ (Samples C and E), the profile exhibits a $V$-shaped feature around $B=0$ which we believe may be related to the long-range magnetic order that sets in below 8 K. The shoulders and local maxima rapidly become pronounced when $\bf B$ is tilted away from the longitudinal direction, as shown in Fig. 2a (main text).

\section{Chiral anomaly and its doping dependence}
\label{sec:doping_mr}
\begin{figure}[htb]
  \centering
  \includegraphics[width=9.2cm]{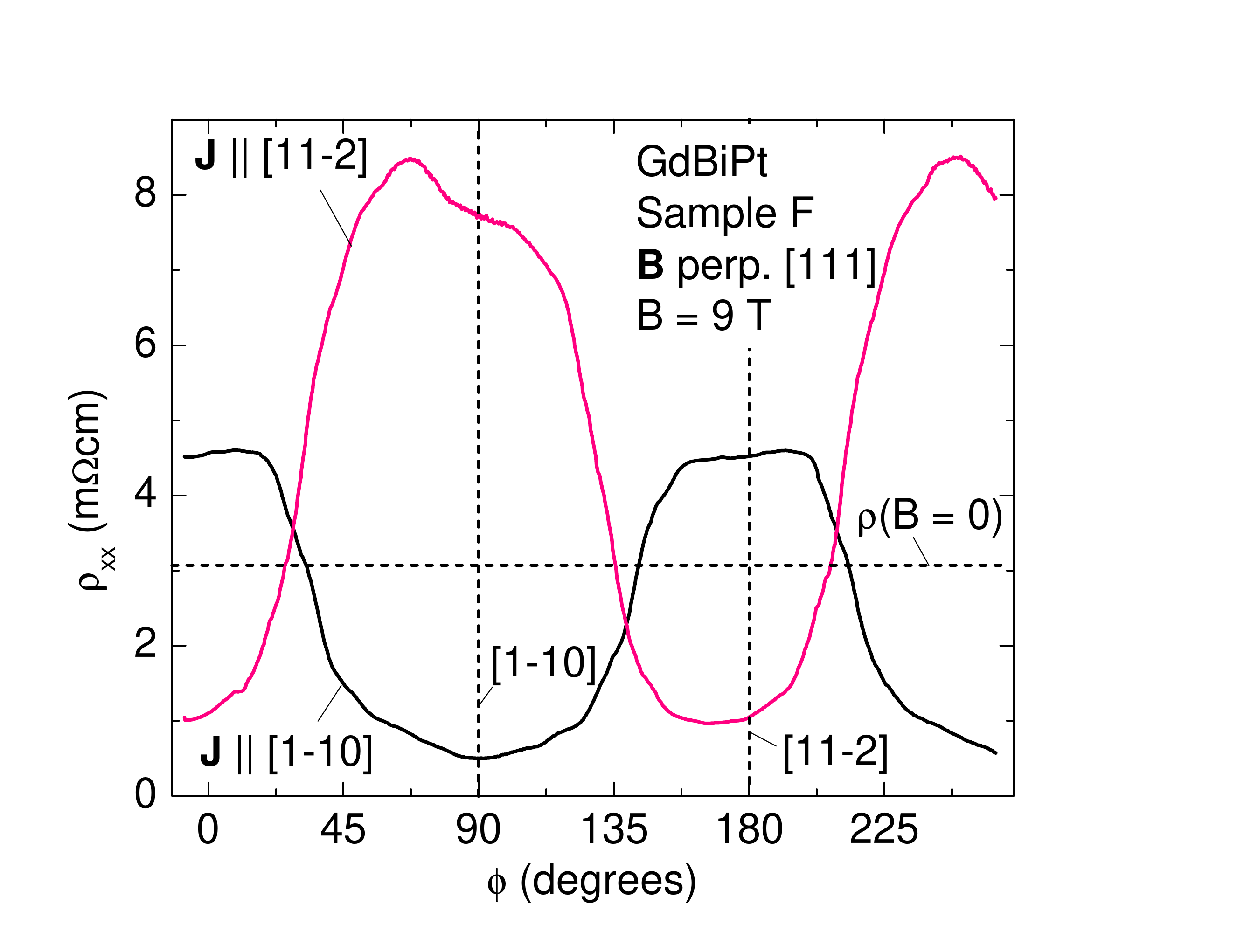}
  \caption{Field steering of the 'axial plume' for sample F. We attached several contacts to the sample so that the direction of the electrical current $\bf J$ could be changed in situ. We observe that the negative magneto-resistance only occurs when $\bf J$ and $\bf B$ are aligned in parallel. The magnetic field was rotated in the in-plane geometry, with $\bf{B} \perp$ [111]. The magnitude of $\bf B$ was $9\,$T in this experiment.}
\label{fig:samplef_rotate}
\end{figure}

\begin{figure}[htb]
  \centering
  \includegraphics[width=9.2cm]{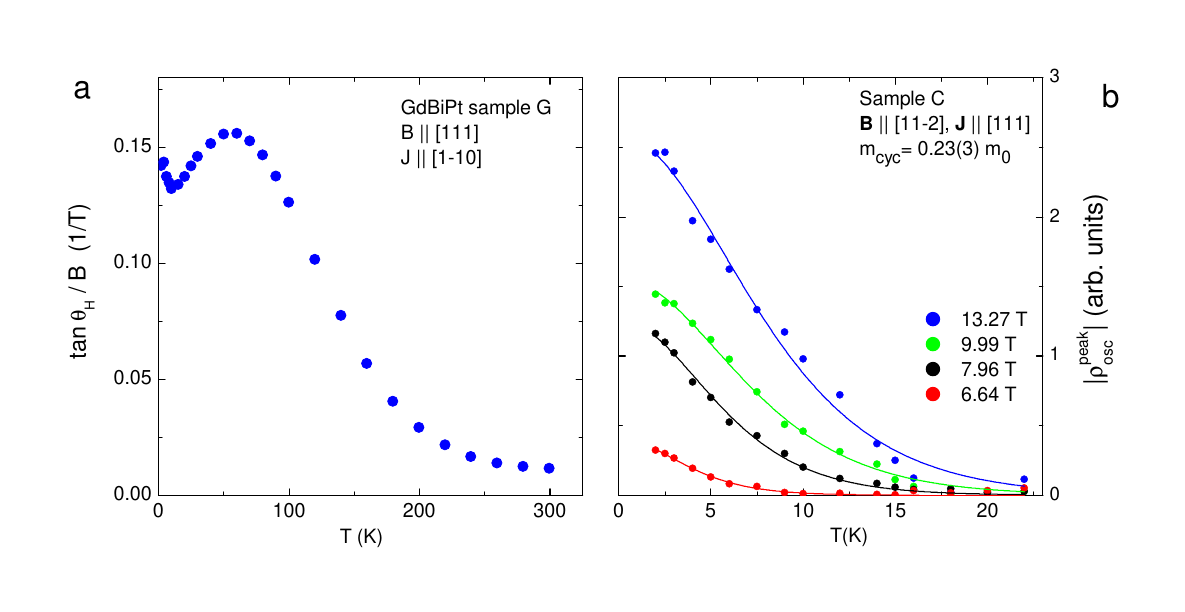}
  \caption{Estimates of the mobility from the Hall angle and of the effective mass from SdH oscillation amplitudes. Panel (a) plots the $T$ dependence of the measured tangent of the Hall angle $\tan\theta_H$ in Sample G. Below 10 K, the value $\tan\theta_H/B\sim$ 0.15 T$^{-1}$ (hole-like) corresponds to a mobility $\mu\sim$ 1,500 cm$^2$/Vs. Above 100 K, copius excitation of $n$-type carriers adds a negative contribution which suppresses the observed Hall angle. Panel (b) plots the SdH amplitudes $\rho^{peak}_{osc}|$ measured at fields from 6.64 to 13.27 T in Sample C. Fits to the Lifshitz-Kosevich expression~\cite{Roth} yield the effective mass $m^*\sim$ 0.23 $m_0$ ($m_0$ is the free electron mass). 
	}
\label{fig:tanHall}
\end{figure}

This section highlights the evolution of the negative longitudinal magnetoresistance (LMR), the established signature of the chiral anomaly in Weyl semimetals \cite{Xiong2015}\cite{Huang2015}, as a function of carrier concentration $n_H$. First, we provide additional data for sample F in Fig. \ref{fig:samplef_rotate}, where the electric current $\bf J$ was applied along two mutually perpendicular crystallographic directions. We observe the negative LMR only when the current and the magnetic field are parallel to each other.\\

Figure \ref{fig:mr_panel} shows the $R$-$T$ curve, magneto-resistance at various angles, and a rotation of the resistance at fixed temperature and magnetic field, for several samples of the same alignment and of varying $n_H$. We track the evolution of $\rho(T)$ from RRR$^{-1}\sim 1$ to insulating behavior, and back to quasi-metallic, in the first column.\\

\begin{figure}[htb]
  \centering
  \includegraphics[width=9.2cm]{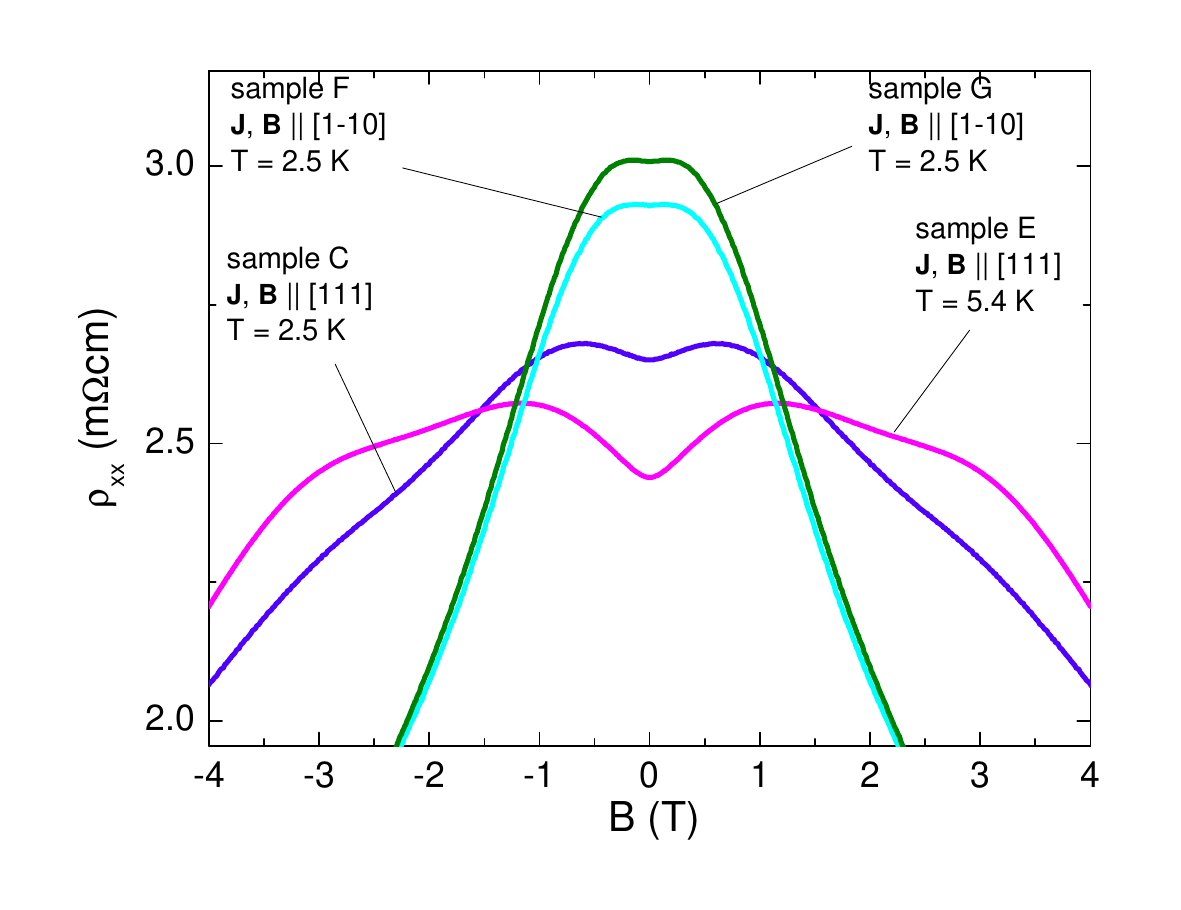}
	\caption{
	The low-field behavior of the LMR in Sample C with $\bf B\parallel$ [111], E ($\bf B\parallel$ [111]), F ($\bf B\parallel [1\bar{1}0]$) and G ($\bf B\parallel [1\bar{1}0]$). 
	}
\label{Zoom}
\end{figure}

\begin{figure*}[htb]
  \centering
  \includegraphics[width=15.4cm]{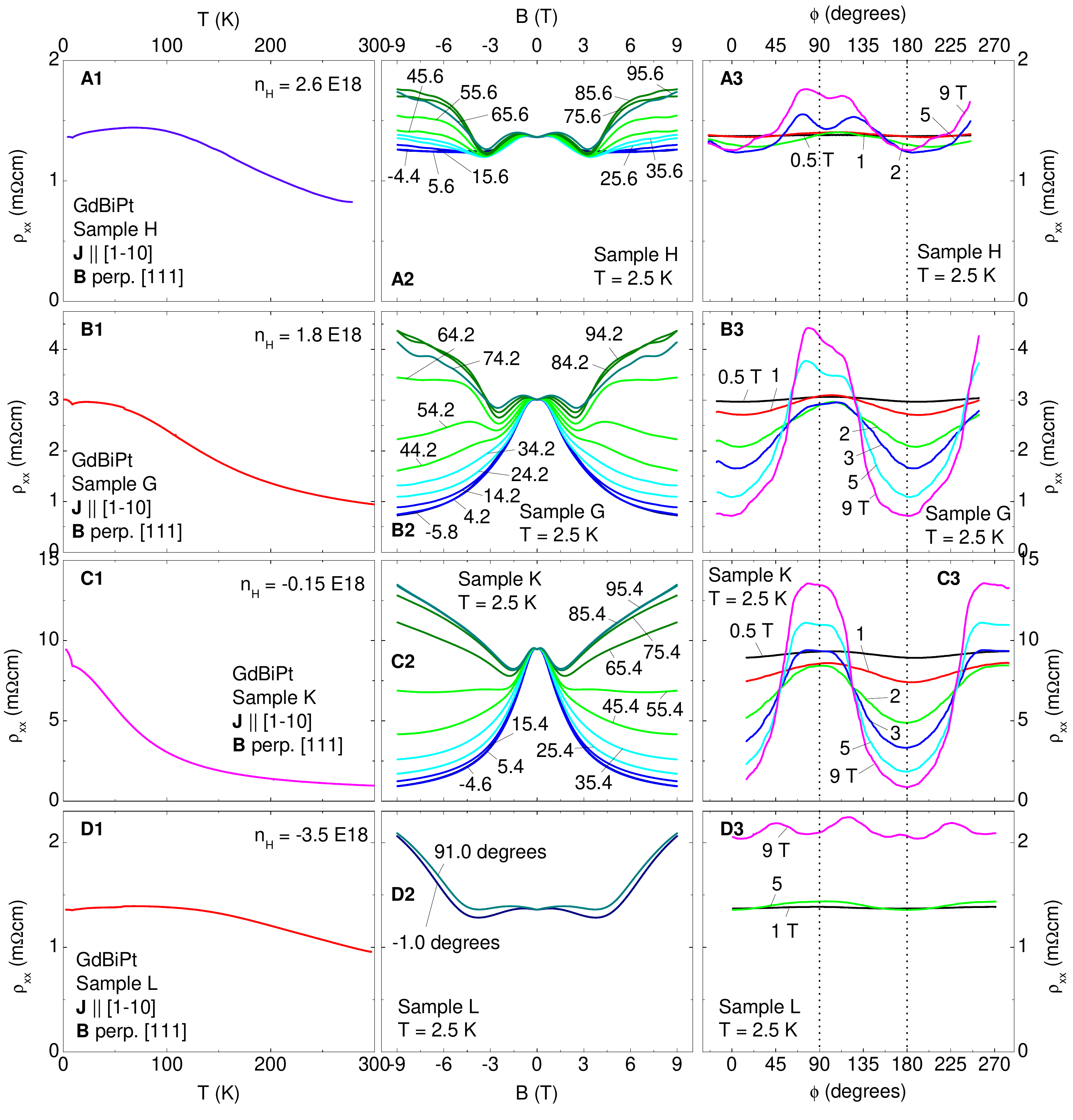}
  \caption{Resistivity $\rho$ vs. temperature $T$ (RT) curves, magneto-resistivity at various angles, and in-plane rotation of $\rho$ for fixed field in samples of various Hall carrier concentrations $n_H$ (in cm$^{-3}$). The current $\bf J$ is applied parallel to the [1-10] axis (or equivalent) in all cases, and the magnetic field is rotated in the plane perpendicular to [111]. See main text for discussion.}
\label{fig:mr_panel}
\end{figure*}

The transverse magneto-resistance at $\phi = 90^\circ$ in the center column shows a negative dip at low field in all samples. This feature may be attributed to the field-induced reconstruction of the electronic structure, or - less likely to us - magnetism. The higher mobility of the light electron band in the inverted band structure of GdPtBi may be the reason why the positive contribution to the transverse MR is more pronounced for sample L (n-type), as compared to sample H (p-type). \\

The LMR is suppressed for the most strongly doped samples (panels A2 and D2). We find the biggest LMR in sample K, which has one of the lowest $\left| n_H \right|$ of all samples studied.\\

Slight misalignments of the current direction with respect to a crystallographic axis become most apparent in the $\rho(\phi)$ plots in the third column of Fig. \ref{fig:mr_panel}. We believe this to be the cause of the anti-symmetries especially apparent in panels A3 and B3. In panel D3, the $9\,$T curve may be understood as a superposition of a six-fold pattern, characteristic of magnetic field aligned in the (111) plane of a cubic lattice, with the two-fold pattern of longitudinal vs. transverse alignment of $j$ and $\textbf{B}$; however, the difference between transverse and longitudinal MR is now very small, at large $\left| n_H\right|$.\\

\begin{figure*}[htb]
  \centering
  \includegraphics[width=15.4cm]{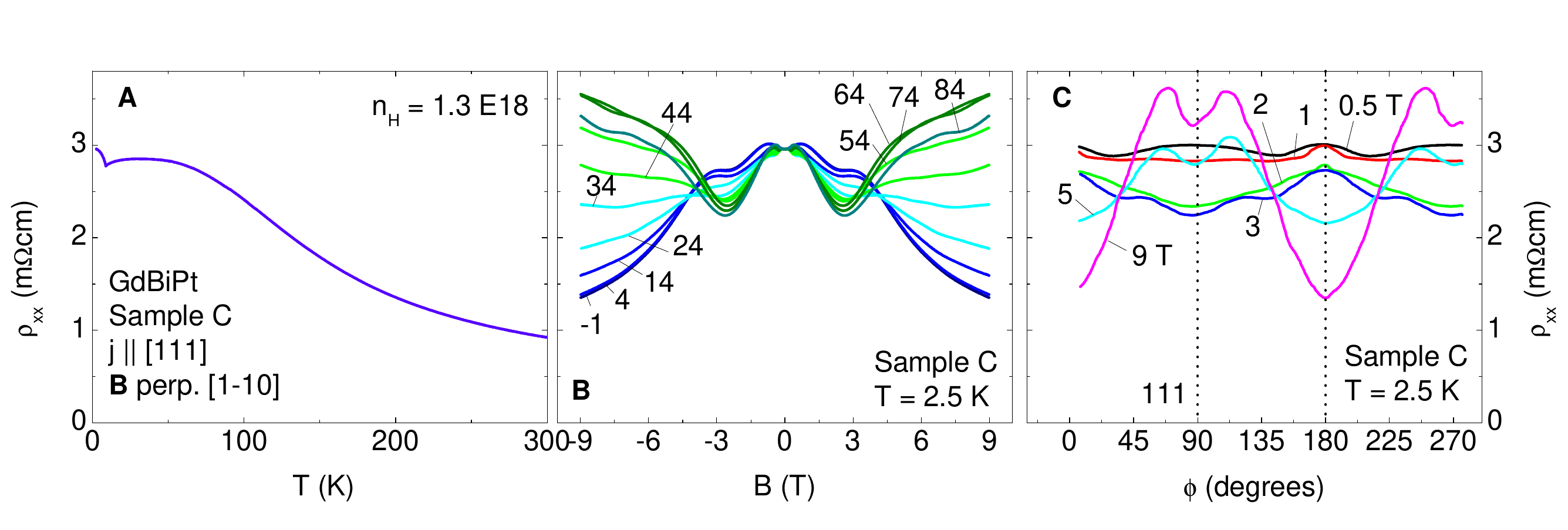}
  \caption{Resistivity $\rho$ vs. temperature $T$ (RT) curve (A), magneto-resistivity at various angles (B), and in-plane rotation of $\rho$ (C) for sample C, where current $\bf J$ is applied parallel to [111], and the field is rotated in the plane perpendicular to [11-2].}
\label{fig:mr_direction}
\end{figure*}

Sample C was prepared as a plate with largest face perpendicular to [11-2], and current $\bf J$ along [111] (Fig. \ref{fig:mr_direction}). Sample G was cut from the same batch and had similar $n_H$, but with ${\bf J} \parallel [1-10]$ (see Fig. \ref{fig:mr_panel} panels B1 - B3). Comparing the two samples which have different directions of $\bf J$, it is clear that the dip in the transverse MR is weaker for magnetic field $\textbf{B}$ in the [111] direction. We also find that the LMR is weaker with current $\bf J$ in the [111] direction. These variations may be a consequence of the anisotropic changes in the electronic structure, when a magnetic field is applied (see e.g. Fig. \ref{fig:dos}).\\

\begin{figure}[htb]
  \centering
  \includegraphics[width=9.2cm]{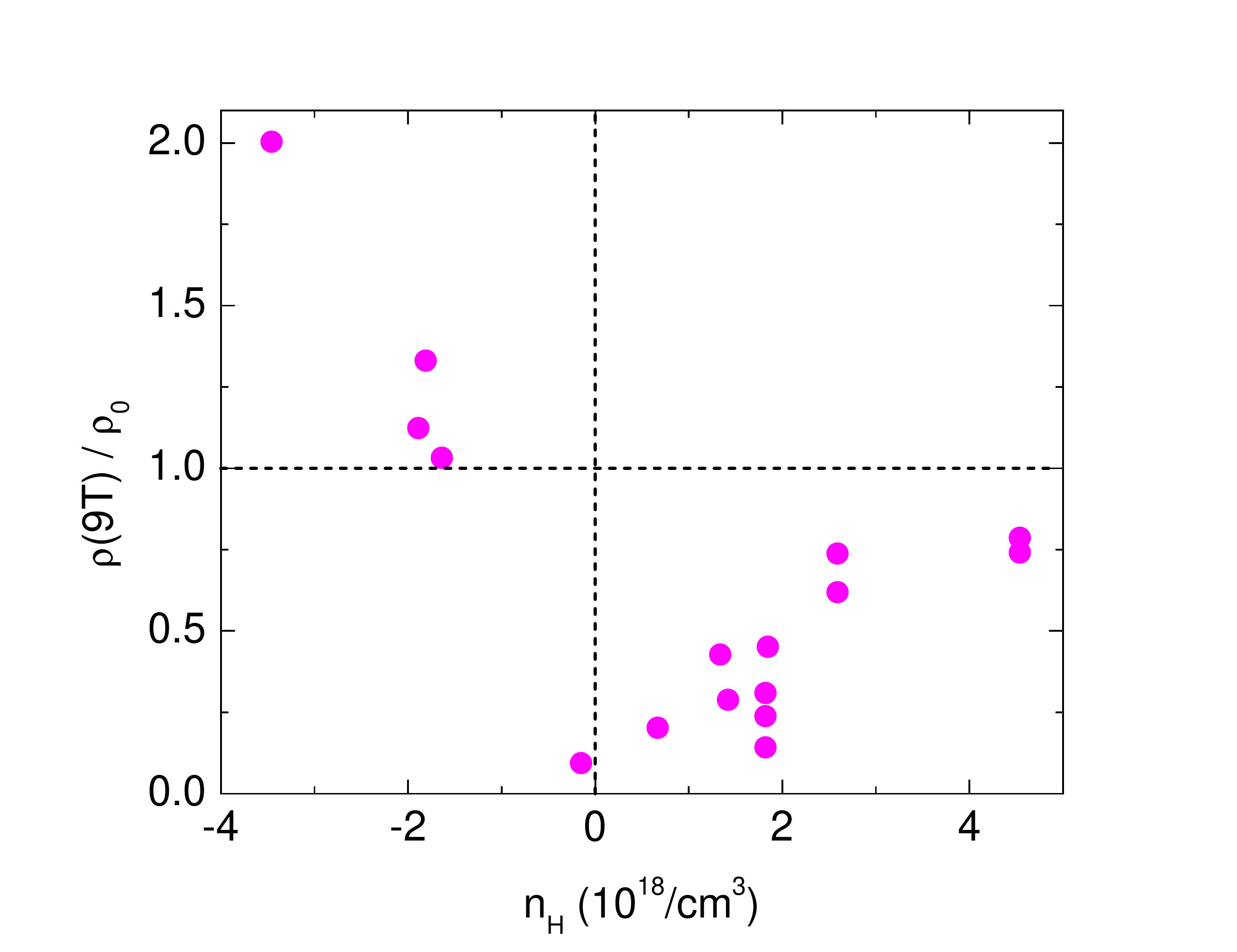}
  \caption{Negative longitudinal magnetoresistance (LMR) as a function of carrier concentration $n_H$ from the Hall effect. The relative suppression of $\rho_{xx}$ in a longitudinal magnetic field is strongest close to $n_H = 0$.}
\label{fig:mrrat_vhalldens}
\end{figure}

The negative LMR may be visualized as a function of carrier density $n_H$ by plotting the relative change $\delta \rho = \rho_\parallel (9\,$T$)/\rho_0$ (Fig. \ref{fig:mrrat_vhalldens}). We observe a strong dependence of $\delta \rho$ on $n_H$, and significant anti-symmetry with respect to negative / positive carrier concentration. This may be at least in part due to the difference in effective mass for the heavy hole and light electron bands (at zero field), where the same $\left| n_H\right|$ corresponds to a larger Fermi energy $\left|E_F\right|$ in the case of the light electrons. To us, Fig. \ref{fig:mrrat_vhalldens} provides firm evidence that the negative LMR we observe is a consequence of the topological nature of the band structure in GdPtBi.\\

\section{Thermal conductivity and thermopower}
\label{sec:kxx}


We have measured the thermal conductivity for several samples of different carrier concentrations $n_H$ and find that $\kappa_{xx}$ is unusually large at low temperature. From Wiedemann-Franz comparison with the conductivity, we infer that $\kappa$ is overwhelmingly dominated by phonon conduction. The large phonon term reflects the stiff moduli of the the half-Heusler lattice. The weak variation of $\kappa_{xx}$ at $300\,$K with doping lies within our margin of error $\sim 20\,$\% due to the measurements of the sample geometry. However, the maximum of $\kappa_{xx}$ at low temperature is largest for samples with small $\left| n_H \right|$ - consistent with the notion that the cleanest crystals have the lowest dopant concentration. The direction of the heat current $\bf J_Q$ for each of these samples is the same as the direction of the electrical current $\bf J$ listed in Table \ref{tab:samples}. The data were taken in zero magnetic field, but we have found no evidence for a change of $\kappa_{xx}$ with $\bf B$ at $T>7\,$K.\\

The sign of the thermopower at low temperature agrees with the sign of the Hall effect, consistent with the single band picture. The amplitude of the thermopower decreases towards high temperatures, where thermally activated carriers play a larger and larger role. The temperature dependence of $S_{xx}/T$ (not shown) gives evidence for saturation at low $T$. We extract an estimate for the Fermi energy $E_F\sim 2.0\,\beta$meV (samples G, E) using the standard Mott formula $S(T) = (\pi^2/3)(k_B/e)(k_B T/E_F)\beta$ and the value of $S_{xx}/T$ at $10\,$K. The dimensionless parameter $\beta$ is the exponent describing how the density of states ${\cal N}$ and velocity vary with $E$ (see e.g. \cite{Liang2013}). Below $10\,$K, the curves of $S_{xx}/T$ are distorted by the magnetic transition at $T_N = 8.8\,$K. The extension of our thermopower experiments to lower temperature would be required to get a better estimate for $E_F$.

\section{Ruling out inhomogeneous current and ``current jetting'' effects}
\label{sec:jetting}

We have performed a series of tests to address the concern that the observed negative LMR may arise from a combination of current inhomogeneities caused by disorder and the effect of current jetting which is important in high mobility samples in LMR experiments.\\

\noindent\emph{Disorder}\\
As mentioned in the main text and in Methods, we tested for inhomogeneous current distribution. Voltage contacts on Sample G were remounted, replacing the previous ones with 10 new, small voltage contact pads. The current contacts A and B are sufficiently large to cover the shorter edges of the crystal. At $T$ = 2 K, we measured simultaneously the potential difference $V_{i-j}$ between the 8 pairs of nearest-neighbor contacts ($V_{1-2}$, $V_{2-3}, \cdots, V_{9-10}$) as $B$ (applied $\parallel {\bf J}\parallel {\bf \hat{x}}$) is varied.

The 8 curves for the relative change in $V_{i-j}$ are nearly identical below 3 T, only displaying slight, non-systematic deviations above 5 T (Fig. 3d of main text). To us, the striking agreement across the 8 contacts provides strong evidence for the uniformity of $\bf J(r)$ throughout the crystal. In Panel (C), we plot the angular variation of the 8 quantities $V_{i-j}$ (expressed as a relative resistance $R(9T)/R(0)$) as $\bf B$ is rotated in the $x$-$y$ plane (with $B$ fixed at 9 T). Again, the 8 curves agree well with each other over the broad angular interval $-60^\circ\to 60^\circ$ ($\bf B\parallel J$ at $\phi = 0$). When $\bf B$ is $\perp\bf J$, however, there exists deviations of up to $\sim 10\%$ between the curves which arise from the transverse MR. Panel (D) shows the 8 transverse MR curves measured with $\bf B\perp J$. The deviations are now larger, with pairs closer to the middle (3-4 and 7-8) showing a larger MR than the ones closer to the current contacts (6-7 and 4-5). [In the transverse geometry with $\bf B\parallel \hat{y}$, the moderately large Hall angle ($\tan\theta_H\sim 2$ at 9 T) now causes significant distortions of the current distribution at the corners of a uniform crystal. We attribute the deviations to the Hall effect. The experiment shows that the distortions are minimal in the LMR geometry.] 

From the 8 curves, we infer that the current density is uniform throughout the crystal in the LMR experiment. This implies that the observed negative LMR is an intrinsic electronic effect rather than arising from strong distortions of the current paths.\\


\noindent\emph{Current Jetting}\\
In materials with high mobility ($\mu\gg$ 50,000 cm$^2$/Vs), negative LMR can arise from current jetting \cite{PippardMagres}\cite{Reed1971}. We take the the sample plane to be normal to $\bf\hat{z}$. In the Drude model, with the magnetic field $\bf B\parallel{\bf J}\parallel\bf\hat{x}$, the longitudinal conductivity is $\sigma_{xx} = \sigma_0$ (a constant). The conductivity transverse to $\bf B$ decreases as $\sigma_{yy} = \sigma_0/[1+(\mu B)^2]$ where $\mu$ is the mobility. In the limit $\mu B\gg 1$, the large anisotropy forces $\bf J$ to flow mostly $\parallel\bf B$ everywhere while flow transverse to $\bf B$ is strongly suppressed. As a result, the potential gradient $\nabla \psi$ becomes very large near the contacts while its magnitude at a sample edge is suppressed significantly. This current jetting effect is most serious when the current sources and sinks are point sources.

\begin{figure*}[htb]
  \centering
	\includegraphics[width=9cm]{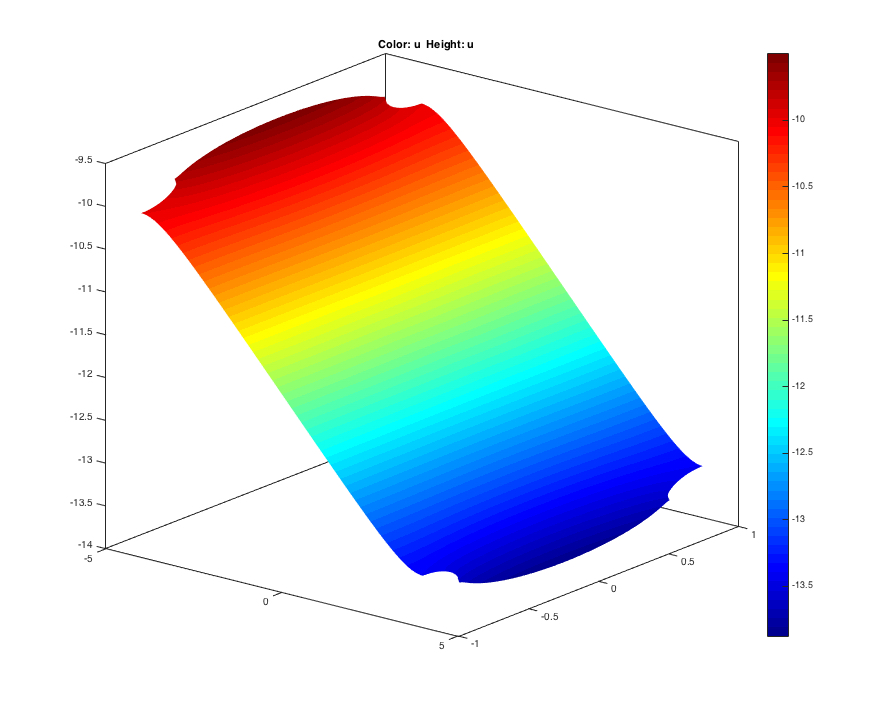}
  \includegraphics[width=9cm]{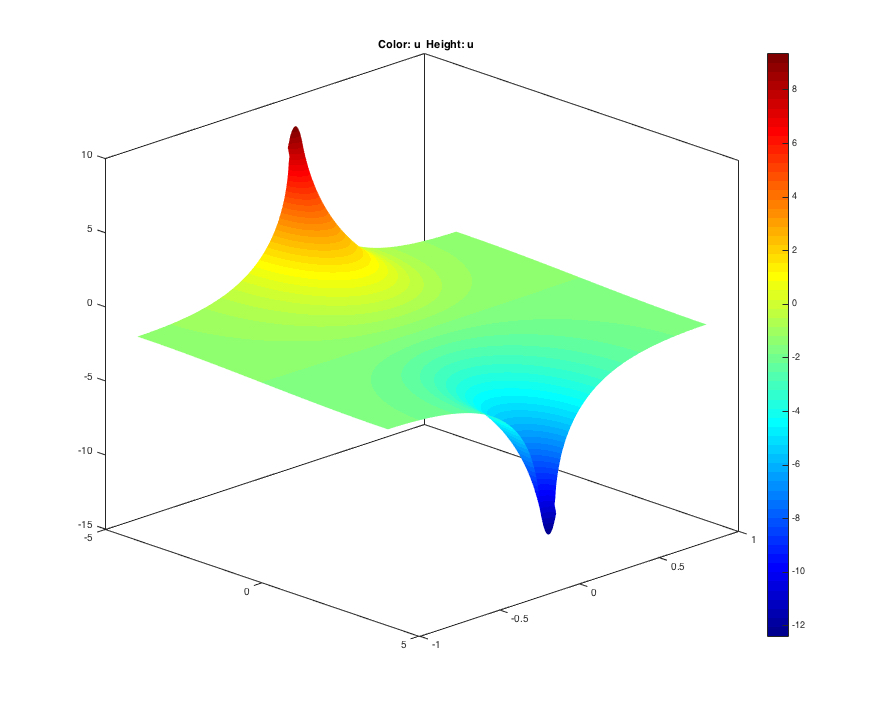}
			\caption{Comparison of the potential function $\psi(x,y)$ in a sample at moderate field ($\mu B$ = 2) and in a very large field ($\mu B$ = 10) with $\bf B$ applied $||{\bf J}$. The sample, of aspect ratio $\ell:w$ = 4:1, has a mobility $\mu$ = 2,000 cm$^2$/Vs. In the upper panel, we assume a large current-contact width ($w_c = 0.75 w$). With $B$ = 10 T, the anisotropy $\sigma_{xx}/\sigma_{yy}$ = 5. The simulated potential displays a gradient $-\nabla \psi$ that is quite uniform everywhere, showing minimal distortions from current jetting. In the lower panel, $B$ is increased to 50 T (anisotropy increased to 100) and $w_c$ reduced to 0.05 $w$. The effects of current jetting now become pronounced. The steep variation of $\psi$ in the vicinity of the contacts leads to strong reduction of $|\nabla\psi|$ at the edges of the sample.
	}
	\label{figjets}
	\end{figure*}

As mentioned in Methods, we have performed extensive numerical simulations to rule out current jetting as the origin of the negative longitudinal MR in GdPtBi. We report the 2D simulations here (a few 3D simulations were performed as well to check that similar results are obtained).
The two equations $\nabla \cdot {\bf J} = 0$ and $J_i = \sigma_{ij}E_j$, where $E_j = -\partial_j\psi$, imply that the potential function $\psi(x,y)$ satisfies the anisotropic 2D Laplace equation
\begin{equation}
[\partial_x\sigma_{xx}\partial_x + \partial_y\sigma_{yy}\partial_y ]\psi(x,y) = 0.\nonumber
\label{Laplace}
\end{equation}
Instead of rescaling the sample dimensions $w$ and $\ell$ (to obtain the isotropic equivalent), we solve Eq. \ref{Laplace} directly using the relaxation method on a 2D triangulated mesh network that covers the sample with a MatLab subroutine. Neumann boundary conditions are imposed, viz. ${\bf \hat{n}}\cdot{\bf J} = \pm c$ (or 0), where $\bf\hat{n}$ is the unit vector normal to the boundary and $c$ is a constant source term. For the thinnest plate-like samples, the 2D model provides useful semiquantitative guidance. 

As the program converges to the final solution the mesh density increases by a factor of 100. We compare in Fig. \ref{figjets} the variation of $\psi(x,y)$ for a case close to our experiment (current contact pads of width $w_c = 0.75w$, $B$= 10 T) with the case when $B$ is increased to 50 T with current contacts reduced to small points ($w_c$ = 0.05$w$). The mobility is held at 2,000 cm$^2$/Vs in both panels. In Fig. \ref{figjets} (upper panel) the nearly uniform gradient $\nabla\psi(x,y)$ implies that $J$ is close to being uniform everywhere. By contrast, in the lower panel, the existence of pronounced jets leads to very large $|\nabla\psi|$ in the vicinity of each current contact, together with strong suppression of $|\nabla \psi|$ at the sample's edges. This illustrates the current jetting effect.

\begin{figure*}[htb]
  \centering
	\includegraphics[width=10cm]{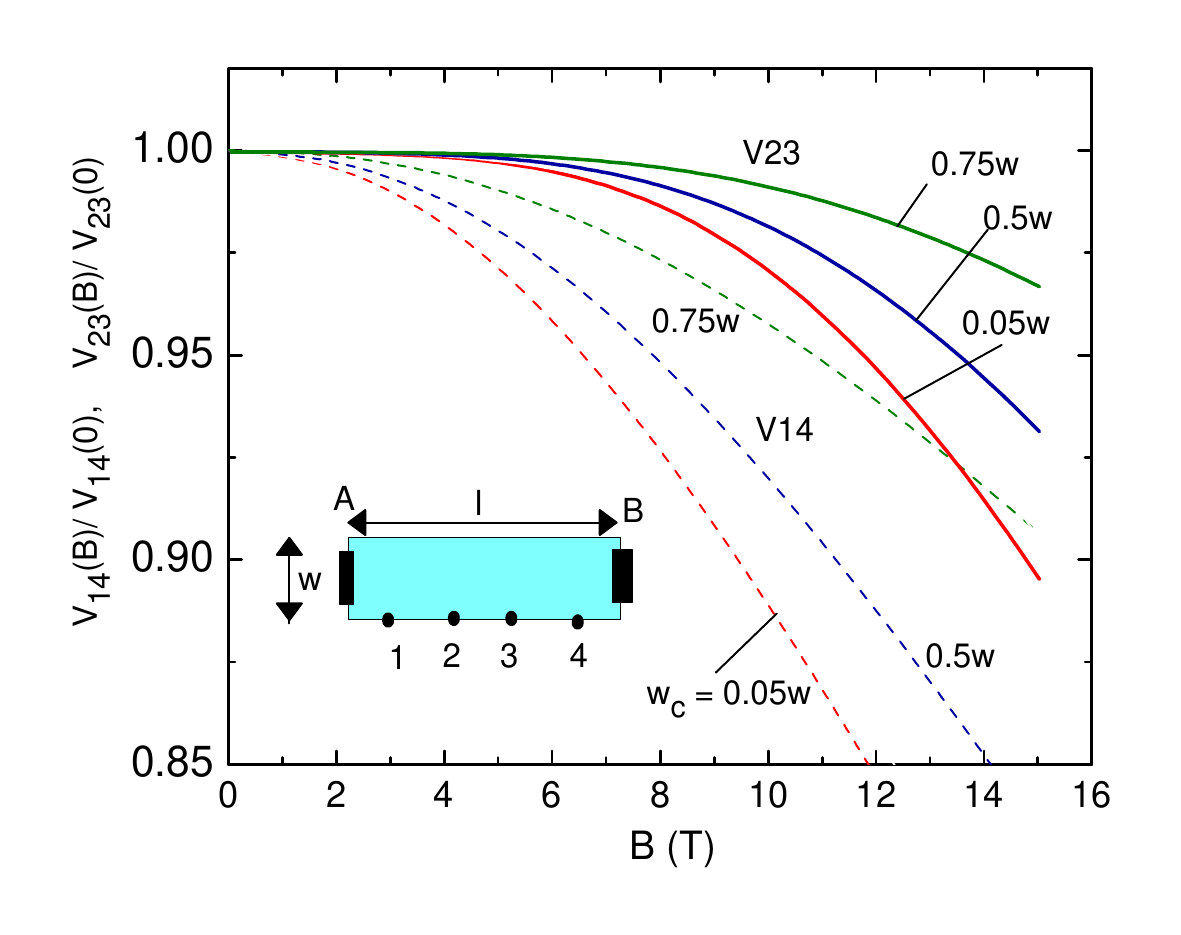}
			\caption{The fractional change in voltage $V_{ij}$ across the contacts $i$ and $j$ versus $B$ calculated numerically in a rectangular sample of width $w$ and length $\ell$ for 3 widths of the current contacts A and B ($w_c$ = 0.05$w$, 0.5$w$ and 0.75$w$). Our experiment is closest to the case $w_c = 0.7w$. The mobility is taken to be 2,000 cm$^2$/Vs. The contacts are labelled in the inset. At $B$ = 10 T, the relative change in $V_{14}$ varies from 4-10 $\%$ (dashed curves). For a contact pair closer to the middle of the sample, $V_{23}$, the relative change is only 1-3 $\%$. These changes are far too small to account for the observed LMR in GdPtBi. For e.g. the change induced by current jetting is $<1\%$ at 2 T whereas the observed $\rho_{xx}$ in Fig. 1d of main text has decreased by 2$\times$.
	}
	\label{figVB}
	\end{figure*}

To quantify the results from the simulation, we have extracted the voltage potential differences $V_{ij}$ read across various contacts $i$ and $j$. Their field profiles are shown in Fig. \ref{figVB} (for a sample similar to G with aspect ratio $\ell/w$ = 4 and a mobility $\mu$ = 2,000 cm$^2$/Vs). The dashed curves represent $V_{14}$ (contacts separated by 0.8 $\ell$) for 3 values of the current-contact widths ($w_c$ = 0.05 $w$, 0.5 $w$ and 0.75 $w$). Even in the most precarious case (point-like contacts), current jetting suppresses the observed $V_{14}$ by only 9$\%$ at our maximum field (9 T). Increasing the contact widths to 0.75 $w$ (closest to our experiment) decreases the change to 4$\%$. The solid curves show the corresponding cases for $V_{23}$ (contacts separated by 0.20 $\ell$ near the middle of the sample). Now the suppression is between 1 and 3$\%$. Hence, current jetting effects are mitigated by increasing the contact pad widths and sampling far from current contact pads. Our experiment is closest to the case with $w_c = 0.75 w$. 

Further simulations (to be reported elsewhere) into optimal experimental configurations reveal that merely adopting the standard Hall-bar configuration for LMR experiments does not guard against current jetting effects.

We conclude that, below 10 T, current jetting has a negligible effect (a few $\%$) in a sample with $\mu$ = 2,000 cm$^2$/Vs. This is far too small to account for the large LMR in GdPtBi. Note that from Fig. 3d of the main text, $\rho_{xx}$ has decreased by 2$\times$ at the low field of 2 T. We would need $B>$50 T to achieve the same suppression if current jetting were the origin.


\end{document}